\documentclass[structabstract]{aa} 
\usepackage{txfonts}
\usepackage{graphicx}
\usepackage[usenames,dvipsnames]{color}
\usepackage{xspace} % allows for ``smart'' trailing spaces (space if space, no space if punctuation).
\usepackage{url}
\usepackage{natbib}
\bibpunct{(}{)}{;}{a}{}{,}

\newcommand{\ms}{\ensuremath{\textrm{m\,s}^{-1}}}
\newcommand{\kms}{\ensuremath{\textrm{km\,s}^{-1}}}

\newcommand{\zem}{\ensuremath{z_{\rm em}}}
\newcommand{\zabs}{\ensuremath{z_{\rm abs}}}

\newcommand{\daa}{\ensuremath{\Delta\alpha/\alpha}}

\newcommand{\ma}{\mbox{m\AA}}
\newcommand{\cm}{cm$^{-2}\,$}

\newcommand{\pixel}{pixel$^{-1}$}

\newcommand{\popler}{\textsc{uves\_popler}}

\newcommand{\LogN}[2]{\ensuremath{\log N}(\ion{#1}{#2})}

\newcommand{\systemARedshift}{1.6919}
\newcommand{\systemBRedshift}{1.6279}
\newcommand{\systemCRedshift}{1.5558}
\newcommand{\systemDRedshift}{0.9424}
\newcommand{\systemERedshift}{0.9407}
\newcommand{\systemFRedshift}{0.7866}

\newcommand{\ReportStatisticalError}[2]{\ensuremath{#1\,\pm\,#2_{\rm stat}}}
\newcommand{\ReportStatisticalandSystematicError}[3]{\ensuremath{#1\,\pm\,#2_{\rm stat}\,\pm\,#3_{\rm sys}}}

\newcommand{\psystemAdaaOnePlace}{-3.8}
\newcommand{\psystemAdaaTwoPlaces}{-3.84}
\newcommand{\psystemAdaaStatisticalErrorOnePlace}{2.1}
\newcommand{\psystemAdaaStatisticalErrorTwoPlaces}{2.10}
\newcommand{\psystemAdaaChiSquare}{0.96}

\newcommand{\psystemAsidamdaaOnePlace}{+1.1}
\newcommand{\psystemAsidamdaaTwoPlaces}{+1.14}
\newcommand{\psystemAsidamdaaStatisticalErrorOnePlace}{2.6}
\newcommand{\psystemAsidamdaaStatisticalErrorTwoPlaces}{2.58}
\newcommand{\psystemAsidamdaaChiSquare}{0.94}

\newcommand{\wsystemAdaaOnePlace}{+1.3}
\newcommand{\wsystemAdaaTwoPlaces}{+1.29}
\newcommand{\wsystemAdaaStatisticalErrorOnePlace}{2.4}
\newcommand{\wsystemAdaaStatisticalErrorTwoPlaces}{2.35}
\newcommand{\wsystemAdaaSystematicErrorOnePlace}{1.0}
\newcommand{\wsystemAdaaSystematicErrorTwoPlaces}{0.96}
\newcommand{\wsystemAdaaChiSquare}{1.19}

\newcommand{\wsystemBdaaOnePlace}{+37.2}
\newcommand{\wsystemBdaaTwoPlaces}{+37.22}
\newcommand{\wsystemBdaaStatisticalErrorOnePlace}{20.6}
\newcommand{\wsystemBdaaStatisticalErrorTwoPlaces}{20.62}
\newcommand{\wsystemBdaaSystematicErrorOnePlace}{39.4}
\newcommand{\wsystemBdaaSystematicErrorTwoPlaces}{39.44}
\newcommand{\wsystemBdaaChiSquare}{1.43}

\newcommand{\wsystemCdaaOnePlace}{-0.2}
\newcommand{\wsystemCdaaTwoPlaces}{-0.15}
\newcommand{\wsystemCdaaStatisticalErrorOnePlace}{24.9}
\newcommand{\wsystemCdaaStatisticalErrorTwoPlaces}{24.90}
\newcommand{\wsystemCdaaSystematicErrorOnePlace}{3.6}
\newcommand{\wsystemCdaaSystematicErrorTwoPlaces}{3.55} 
\newcommand{\wsystemCdaaChiSquare}{1.21}

\newcommand{\wsystemDdaaOnePlace}{+9.3}
\newcommand{\wsystemDdaaTwoPlaces}{+9.32}
\newcommand{\wsystemDdaaStatisticalErrorOnePlace}{12.7}
\newcommand{\wsystemDdaaStatisticalErrorTwoPlaces}{12.68}
\newcommand{\wsystemDdaaSystematicErrorOnePlace}{9.8}
\newcommand{\wsystemDdaaSystematicErrorTwoPlaces}{9.84} 
\newcommand{\wsystemDdaaChiSquare}{1.30}

\newcommand{\wsystemEdaaOnePlace}{+21.4}
\newcommand{\wsystemEdaaTwoPlaces}{+21.40}
\newcommand{\wsystemEdaaStatisticalErrorOnePlace}{24.2}
\newcommand{\wsystemEdaaStatisticalErrorTwoPlaces}{24.16}
\newcommand{\wsystemEdaaSystematicErrorOnePlace}{20.9}
\newcommand{\wsystemEdaaSystematicErrorTwoPlaces}{20.86}
\newcommand{\wsystemEdaaChiSquare}{1.13}

\newcommand{\wsystemFdaaOnePlace}{+7.3}
\newcommand{\wsystemFdaaTwoPlaces}{+7.32}
\newcommand{\wsystemFdaaStatisticalErrorOnePlace}{35.9}
\newcommand{\wsystemFdaaStatisticalErrorTwoPlaces}{35.89}
\newcommand{\wsystemFdaaSystematicErrorOnePlace}{21.4}
\newcommand{\wsystemFdaaSystematicErrorTwoPlaces}{21.36} 
\newcommand{\wsystemFdaaChiSquare}{1.29}

\begin{document}

\title{The UVES Large Program for Testing Fundamental Physics: I Bounds on a change in \boldmath{$\alpha$} towards quasar HE\,2217$-$2818 \thanks{Based on observations taken at ESO Paranal Observatory. Program L 185.A-0745}}
%\subtitle{expectations, limitations and status quo}

\author{P. Molaro\inst{1,7}, M. Centurion\inst{1}, J. B. Whitmore\inst{2}, T. M. Evans \inst{2}, M. T. Murphy\inst{2}, I. I. Agafonova \inst{3}, P. Bonifacio\inst{4}, S. D'Odorico\inst{5}, S. A. Levshakov\inst{3,12}, S. Lopez\inst{6}, C. J. A. P. Martins\inst{7}, P. Petitjean\inst{8}, H. Rahmani\inst{10}, D. Reimers\inst{9}, R. Srianand\inst{10}, G. Vladilo\inst{1}, M. Wendt\inst{11,9}
}
\offprints{P. Molaro}

\institute{ INAF-Osservatorio Astronomico di Trieste, Via G.\,B.\,Tiepolo 11,
34131 Trieste, Italy\\
\email{molaro@oats.inaf.it}
\and Centre for Astrophysics and Supercomputing, Swinburne University of Technology, Hawthorn, VIC 3122, Australia
\and Ioffe Physical-Technical Institute, Polytekhnicheskaya, Str. 26, 194021 Saint Petersburg, Russia
\and GEPI, Observatoire de Paris, CNRS, Univ. Paris Diderot, Place Jules Janssen, 92190 Meudon, France
\and ESO Karl, Schwarzschild-Str. 1 85741 Garching, Germany
\and Departamento de Astronomia, Universidad de Chile, Casilla 36-D, Santiago, Chile
\and Centro de Astrof\'isica, Universidade do Porto, Rua das Estrelas, 4150-762 Porto, Portugal
\and Universite Paris 6, Institut d'Astrophysique de Paris, CNRS UMR 7095, 98bis bd Arago, 75014 Paris, France
\and Hamburger Sternwarte, Universit\"at Hamburg, Gojenbergsweg 112, 21029 Hamburg, Germany
\and Inter-University Centre for Astronomy and Astrophysics, Post Bag 4, Ganeshkhind, 411 007 Pune, India 
\and Institut f\"ur Physik und Astronomie, Universit\"at Potsdam, 14476 Golm, Germany
\and St.~Petersburg Electrotechnical University `LETI', Prof. Popov Str. 5,
197376 St.~Petersburg, Russia
}

\authorrunning{P. Molaro}

\date{Received \today; accepted \today}
\titlerunning{The fine-structure constant towards quasar HE\,2217$-$2818}
%title,aims,methods,results
\abstract {Absorption-line systems detected in quasar spectra can be
  used to compare the value of the fine-structure constant, $\alpha$,
  measured today on Earth with its value in distant galaxies. In
  recent years, some evidence has emerged of small temporal and also
  spatial variations of $\alpha$ on cosmological scales which may
  reach a fractional level of $\approx$10 ppm (parts per million). }
%aim
{ To test  these claims  we are conducting a ``Large Program'' of
  observations with the Very Large Telescope's Ultraviolet and Visual
  Echelle Spectrograph (UVES). We are obtaining high-resolution
  ($R\approx60000$) and high signal-to-noise ratio (${\rm
    S/N}\approx100$) UVES spectra calibrated specifically for this
  purpose. Here we analyse the first complete quasar spectrum from
  this Program, that of HE\,2217$-$2818.}
%methods
{ We apply the Many Multiplet method to measure $\alpha$ in 5 absorption
  systems towards this quasar: $\zabs=\systemFRedshift$, \systemDRedshift,
  \systemCRedshift, \systemBRedshift\ and \systemARedshift.
}
%results
{ The most precise result is obtained for the absorber at
  $\zabs=\systemARedshift$ where 3 \ion{Fe}{ii} transitions and
  \ion{Al}{ii} $\lambda$1670 have high S/N and provide a wide range of
  sensitivities to $\alpha$. The absorption profile is complex, with
  several very narrow features, and requires 32 velocity components to
  be fitted to the data. We also conducted a range of tests to
  estimate the systematic error budget. Our final result for the
  relative variation in $\alpha$ in this system is
  $\daa=\ReportStatisticalandSystematicError{\wsystemAdaaOnePlace}{\wsystemAdaaStatisticalErrorOnePlace}{\wsystemAdaaSystematicErrorOnePlace}$\,ppm. This
  is one of the tightest current bounds on $\alpha$-variation from an
  individual absorber. A second, separate approach to the
  data-reduction, calibration and analysis of this system yielded a
  slightly different result of
  $\ReportStatisticalError{\psystemAdaaOnePlace}{\psystemAdaaStatisticalErrorOnePlace}$\,ppm,
  possibly suggesting a larger systematic error component than our
  tests indicated. This approach used an additional 3 \ion{Fe}{ii}
  transitions, parts of which were masked due to contamination by
  telluric features. Restricting this analysis to the \ion{Fe}{ii}
  transitions only and using a modified absorption profile model, gave
  a result consistent with the first approach,
  $\daa=\ReportStatisticalError{\psystemAsidamdaaOnePlace}{\psystemAsidamdaaStatisticalErrorOnePlace}$\,ppm.
  The other 4 absorbers have simpler absorption profiles, with fewer
  and broader features, and offer transitions with a smaller range of
  sensitivities to $\alpha$. Therefore, they provide looser bounds on
  \daa\ at the $\ga$10\,ppm precision level.}
%conclusion
{ The absorbers towards quasar HE\,2217$-$2818 reveal no evidence for
  variation in $\alpha$ at the 3-ppm precision level (1-$\sigma$
  confidence). If the recently-reported 10-ppm dipolar variation of
  $\alpha$ across the sky were correct, the expectation at this sky
  position is $(3.2$--$5.4)\pm1.7$\,ppm depending on dipole model
  used. Our constraint of
  $\daa=\ReportStatisticalandSystematicError{\wsystemAdaaOnePlace}{\wsystemAdaaStatisticalErrorOnePlace}{\wsystemAdaaSystematicErrorOnePlace}$\,ppm
  is not inconsistent with this expectation.}


\maketitle

\section{Introduction}

Metal lines of absorption systems due to intervening galaxies along
the line of sight towards distant quasars provide insights into the
atomic structure at the cosmic time and location of the intervening
object.  All atomic transitions depend on the fine-structure constant,
offering a way to probe possible variations of its value in space and
time.  The seminal papers of \citet{Savedoff:1956:688} and \citet{Bahcall:1967:L11} took
advantage of the dependence of the relative separation of
fine-structure doublet transitions on $\alpha$ and used the alkali
doublets observed in the first extragalactic sources to limit \daa\ at
the level of a few percent.  More recently, the many-multiplet (MM)
method has been introduced, which allows all observed transitions to
be compared, gaining access to the typically much larger dependence of
the ground state energy levels on $\alpha$ \citep{Dzuba:1999:888}.  Overall,
the MM method improves the sensitivity to the measurement of a
variation of $\alpha$ by more than an order of magnitude over the
alkali-doublet method.  Moreover, the individual dependence on
$\alpha$ of different atoms and transitions allows better control of
instrumental and astrophysical systematic errors.

% For instance $s$--$p$ and $s$--$d$ transitions shift with opposite signs with respect to a change of $\alpha$.
% These transitions allow a direct measure of \daa\ by using the same ion which minimize the impact of ionization effects \citep{Levshakov:2005:827}.

A first analysis using the many-multiplet method on quasar absorption
spectra obtained at the Keck telescope revealed hints that in the
early universe the fine structure constant was smaller than today by
$\sim$6 parts-per-million \citep[ppm;][]{Webb:1999:884,Murphy:2001:1208,Murphy:2003:609}.  However,
the analysis of smaller data sets obtained from the Very Large
Telescope (VLT) in Chile by other groups did not confirm the variation
of $\alpha$ \citep{Quast:2004:L7,Chand:2004:853,Levshakov:2005:827,Levshakov:2006:L21}, though in some cases
these analysis and uncertainty estimates have been challenged  
\citep{Murphy:2007:239001,Murphy:2008:1053}    \citep{Srianand:2007:9002}. A recent analysis has been performed on a
new sample built up by merging 141 measurements from Keck with 154
measurements from VLT \citep{Webb:2011:191101,King:2012:3370}.  The merged sample
indicates a spatial variation in $\alpha$ across the sky at the
4.1$\sigma$ level with amplitude $\sim$10\,ppm.  Since the Keck
telescope in Hawaii is at a latitude of 20$^\circ$~N, and the VLT in
Chile is at latitude of $25^\circ$~S, the two sub-samples survey
relatively different hemispheres and the study of the merged sample
provides a much more complete sky coverage.  Remarkably, the two sets
of measurements are consistent along the region of the sky covered by
the two telescopes.  So far, this spatial variation is formally
consistent with the null results obtained by other groups since the
lines-of-sight fall into a region with minimal reported variation.
%\citep{quast04aap,levshakov05aap,levshakov06aap,levshakov07aap}

It has to be noted that a large-scale spatial variation of this
magnitude is not expected. It cannot be obtained easily in simplest
theoretical models where varying fundamental couplings are due to
dynamical scalar fields.  One can nevertheless reproduce it by
constructing toy models, for example as the result of domains formed
at spontaneous breaking of a discrete symmetry of a dilaton-like
scalar field coupled to electromagnetism \citep{Olive:2011:043509}.  In any case,
if such a result is confirmed, it will highlight the presence of
currently unknown mechanisms which cause nature's physical couplings
to have different values in different regions of the
universe. Observational and experimental confirmation or refutation of
these results is therefore important. In addition to their intrinsic interest, these measurements (whether they are detections or null results) can shed light on the enigma of dark energy  \citep{Amendola:2012:063515}.

%It has to be noted that a spatial variation on large scale is rather unexpected and there are no theories at hand which could provide an explanation. Probably the most pregnant implication could be that this change shows that the values of a fundamental constant are not universal or fixed but the constant may take different values still preserving the fine tuning required by the anthropic principle. It was also suggested that two values of the fine-structure constant could arise from a spontaneous breaking of the symmetry of a dilaton-like scalar field coupled to electromagnetism \citep{Olive:2011:043509}.

One possible weakness in the evidence for variations in $\alpha$ is
that it derives from mostly `archival' Keck and VLT spectra. In
particular, the majority of the spectra analysed by \citet{King:2012:3370} were
obtained in `service' observing mode at the VLT. While spectroscopic
quasar observations are generally straight-forward, the `service'
calibration plan for most VLT/UVES observations only include
wavelength calibration exposures of a thorium--argon (ThAr) lamp many
hours after the relevant quasar exposures. This introduces the
possibility of systematic mis-calibrations of the quasar wavelength
scales. With this possibility in mind, and motivated by the need to
confirm or refute current evidence for variations in $\alpha$, we are
conducting ESO Large Program 185.A-0745, ``{\it The UVES Large Program
  for testing Fundamental Physics}''. The aim is to obtain high
signal-to-noise ratio (${\rm S/N}\sim100$ per pixel) spectra of
$\sim$12 bright quasars, carefully selected to contain a total of
$\sim$25 absorbers along their sight-lines in which \daa\ can be
determined with precision better than $\sim$10\,ppm. In each quasar
line-of-sight there exists at least one absorber in which we expect a
precision approaching $\sim$2\,ppm based on the variety of metal-line
transitions available and the shape/structure of their absorption
profiles.

In the present work we focus on the first line-of-sight from the Large
Program: the bright quasar HE\,2217$-$2818. This sight-line shows 5
separate absorption systems at redshifts 0.9--1.7 in which
$\alpha$-sensitive transitions can be observed at high resolution with
unusually high S/N. This line of sight subtends an angle of
$\Theta\approx58^\circ$ with respect to the pole of the simplest model
of dipolar variation in $\alpha$ across the sky in \citet{King:2012:3370}. The
expected signal at this position on the sky is therefore
$\daa\approx+5.4$\,ppm. Achieving a statistical precision in $\daa$ of
$\la$2\,ppm, as we aim to do in the $\zabs=\systemARedshift$ absorber towards
HE\,2217$-$2818 in this paper using our new, high-S/N observations,
therefore contains a possibility of evidence for or against the
dipolar model of spatial variation in $\alpha$. However, another major
goal of studying this first quasar from the Large Program is to reveal
systematic errors which may not randomize over the larger sample we
will present in future papers. In particular, for the absorption
system at $\zabs=\systemARedshift$, we present the results of two separate
analyses to assist in the assessment of systematic effects.

\section{Observations, data reduction and calibration}

\subsection{Observations}

HE\,2217$-$2818 is a relatively bright quasar with V $\approx$16.0 and
with $\zem = 2.46$.  Its spectrum shows several intervening systems
at $\zabs = 0.600$, \systemFRedshift, \systemDRedshift, 1.054, 1.083, 1.200, \systemCRedshift, \systemBRedshift,
\systemARedshift\ and 2.181.  However, only some of them, i.e.~those detected in
multiple transitions of the low-ionization ions \ion{Fe}{ii},
\ion{Si}{ii} and \ion{Mg}{ii} at \zabs = \systemFRedshift, \systemDRedshift, \systemCRedshift,
\systemBRedshift, and \systemARedshift, are suitable to probe the fine-structure constant.
The other systems have metal lines which are either too weak or which
fall in the Lyman forest short-ward of $\approx$420 nm.  
%The hydrogen
%column densities were not studied because the H{\sc \,i} Ly$\alpha$
%transition of these absorbers falls close to or below the UV
%atmospheric cut off.
%but they are likely Lyman Limit Systems with N(HI)$ \le 10^{18}$ \cm. 

The observations of HE\,2217$-$2818 were performed with UVES on VLT on
the nights of 13, 15, 16 and 17 June 2010, during the first
visitor-mode run of observations for our Large Program.  The journal
of observations and relevant additional information is given in
Table~\ref{table:observations}.
% , more data are included in the table in the Appendix.
The observations comprised 16 separate UVES exposures of
$\approx$4000\,s duration taken on five successive nights for a total
of $\approx$64000 sec of integration.  Half of the observations were
taken with dichroic N.~2 with central wavelengths of the blue and red
arm at 437 and 760 nm (setting 437+760).  The other half were taken
with the dichroic N.~1 in the 390+580\,nm setting.  These two
wavelength settings overlap somewhat, and as a whole they provide
almost complete spectral coverage from 350 up to 947 nm.  The CCDs
were set with no on-chip binning and the pixel size ranged between
0.013-0.015 \AA\ \pixel, or 1.12 \kms \pixel at 400 nm, along the
dispersion direction.  For all observations the slit-width was set to
0.8\arcsec\, providing an instrumental line-spread function with a
FWHM of $\approx$3.57 pixel in the 390 nm frames and $\approx$4.38
pixel in the 580 nm frame, with a variation of about $20\%$ along the
individual echelle orders.  This corresponds to a resolving power of
$R \approx 55000$. During the observations the seeing varied in the
range between 0.5\arcsec\ and 1\arcsec\ as measured by the
Differential Image Motion Monitor telescopes at Paranal
Observatory. However, at the slightly higher and more protected
location of VLT Unit Telescope 2 (on which UVES is mounted) the seeing
was generally 10--30\% better, as measured on the telescope guiding
instrumentation.

%\begin{figure}[]
%\resizebox{\hsize}{!}{\includegraphics[clip=true]{fig_390_fwhm.pdf}}
%\caption{
%\footnotesize Setting 390.
% }
%\label{fig:1}
%\end{figure}

%\begin{figure}[]
%\resizebox{\hsize}{!}{\includegraphics[clip=true]{fig_fwhm_pixels.pdf}}
%\caption{
%\footnotesize FWHM in pixels of the Th-Ar lines across all the orders on top of each other of the 580 nm red low CCD. The figure shows the 
%variation of resolution within the orders. 
% }
%\label{fig:fwhm}
%\end{figure}

% Table observations
% \input{observations} 
 \begin{table}
\caption{Journal of the  observations  of HE 2217-2818. }             
\label{table:observations}      
\centering          
\begin{tabular}{c c c r r  r }     % 7 columns 
\hline
                      % To combine 4 columns into a single one 
&Date  & UT   & $\lambda_{c}(nm)$ & Exp(sec) &     \\ 
\hline

                     &  2010-06-13       & 05:39:48   & 390  & 4000          &      \\
 
                     &  2010-06-13       & 06:51:14   & 390  & 4000          &    \\               
  
                     &  2010-06-13       & 08:03:40   & 390  & 4000          &      \\  
                      &  2010-06-13       & 09:13:00   & 390  & 3600           &      \\      
%        ThAr           &  2010-06-13       & 10:14:08   & 390 &      25            &   10.7  &   10.7   &11.6  &   11.6   &742.95  & 742.95   &  741.5  &  741.5     &1093943  &   \\
%\hline 
%        ThAr          &  2010-06-15       & 05:57:05   & 390  &       25           &    9.4  &   9.4   &10.3 &  10.3   &743.32   &743.32   &  743.0   & 743.0         &1093935   &   \\      
                    &  2010-06-15       & 05:58:25   & 390    & 4000          &     \\
%      ThAr            &  2010-06-15       & 07:06:12   & 390 &      25            &   9.2  &   9.3   &10.1   & 10.1    &743.47  & 743.47  &  742.6 &  742.6           &1093935  &   \\   
                    &  2010-06-15       & 07:07:47   & 390  & 3767           &      \\    
%      ThAr            &  2010-06-15       & 08:11:41   & 390 &      25            &   9.2  &   9.2   &10.0   & 10.0    &743.08  & 743.08  &  742.2 &  742.2           &1093934  &   \\ 
%\hline
%       ThAr           &  2010-06-15       & 08:14:40   & 390 &      25            &   9.2 &   9.2  & 10.0  & 10.0   & 743.08   & 743.08   &  742.2   &  742.2         &1093934  &   \\   
                    &  2010-06-15       & 08:16:02   & 390  & 3767          &    \\  
%   ThAr               &  2010-06-15       & 09:19:56   & 390 &      25            &   9.1  &   9.1   & 9.9 &   9.9   & 742.98   & 742.95  &  742.2 &  742.2              &1093934  &   \\ 
                   &  2010-06-15       & 09:21:48   & 390  & 3767        &    \\                  
%    ThAr             &  2010-06-15      & 10:25:42   & 390 &      25            &   9.0  &   9.0 &  9.8 &   9.8   &743.08  & 743.08   &  742.2 &  742.2                &1093933  &   \\     
%\hline 
%      ThAr           &  2010-06-16       & 05:25:57   & 437 &      25            &   8.5 &  8.5 & 9.2  &9.2  & 744.22   & 744.22   &  743.2         &  743.2          &  1079342  &   \\   
                   &  2010-06-16       & 05:27:19   & 437  & 4020           &    \\  
%  ThAr               &  2010-06-16       & 06:35:26 & 437 &      25            &     8.3  &   8.3   & 9.0 &   9.0   & 743.88   & 743.88 &  742.9 &  742.9             & 1079342  &   \\ 
                   &  2010-06-16       & 06:37:00   & 437  & 4020          &     \\  
%  ThAr               &  2010-06-16       & 07:45:07 & 437 &      25            &     8.2  &   8.2   & 8.8 &   8.9   & 743.77   & 743.77 &  742.8 &  742.8             & 1079342&   \\ 
%\hline
%  ThAr               &  2010-06-16       & 07:48:27 & 437 &      25            &     8.2  &   8.2   & 8.8 &   8.9   & 743.73   & 743.70 &  742.8 &  742.8             & 1079340&   \\  
                   &  2010-06-16       & 07:49:49  & 437  & 4000           &    \\  
%  ThAr               &  2010-06-16       & 08:57:36 & 437 &      25           &     8.0  &   8.0  & 8.7 &   8.7   & 743.48  & 743.50 &  742.5 &  742.5             & 1079340&   \\  
                   &  2010-06-16       & 08:59:10  & 437  & 4020           &       \\  
%ThAr                &  2010-06-16       & 10:07:17 & 437 &      25            &     7.9  &   7.9  & 8.6 &   8.6   & 743.40  & 743.40 &  742.4 &  742.4             & 1079339&   \\  
%\hline 

%  ThAr               &  2010-06-17      & 05:22:41 & 437 &      25            &     8.4  &   8.4   & 9.4 &   9.4   & 743.38   & 743.38 &  742.0 &  742.0            & 1079341&   \\  
                   &  2010-06-17       & 05:27:58  & 437  & 4000           &     \\
                   &  2010-06-17       & 06:35:29  & 437  & 4000           &    \\
%\hline
                   &  2010-06-17       & 07:56:57  & 437  & 4000           &     \\
                   &  2010-06-17       & 09:04:51  & 437  & 4000           &     \\
%ThAr                &  2010-06-17      & 10:12:38 & 437 &      25            &     8.3 &   8.3  & 9.2 &   9.2     & 742.15    & 742.15   &  740.9   &  740.9         & 1079341&   \\ 
% \hline      

%   ThAr                &  2010-06-13       & 05:38:24   & 580  &      5            &    10.8  &   10.8    &11.9 &   11.9    &744.82   & 744.82  &  743.9 &  743.9   & 4690963  &    \\ 
                     &  2010-06-13       & 05:39:44   & 580  & 4000           &      \\
%        ThAr           &  2010-06-13       & 06:49:36   & 580 &      5            &   10.7 &   10.7   &11.8   &   11.8    &743.98  & 743.98 &  742.8 &  742.8  &4690963  &   \\  
                     &  2010-06-13       & 06:51:10   & 580  & 4000           &     \\               
%        ThAr           &  2010-06-13       & 07:58:58   & 580 &       5            &   10.7  &   10.7   &11.7  &   11.7   &743.60    & 743.57   &  742.3   &  742.4    &4690960  &   \\   
%\hline
%        ThAr           &  2010-06-13       & 08:02:12   & 580 &       5            &   10.6 &   10.7 &11.7  &   11.7   &743.52   & 743.52   &  742.3   &  742.3    &4690960  &   \\   
                     &  2010-06-13       & 08:03:36   & 580  & 4000           &       \\  
%        ThAr           &  2010-06-13       & 09:11:24   & 580 &       5            &   10.6  &   10.6   &11.7  &   11.7   &743.22   & 743.22   &  741.9  &  741.9     &4690960  &   \\
                     &  2010-06-13       & 09:12:57   & 580  & 3600  &     \\      
%        ThAr           &  2010-06-13       & 10:14:05   & 580 &       5            &   10.5  &   10.5   &11.6  &   11.6   &742.95  & 742.95   &  741.5  &  741.5     &4690958  &   \\
%\hline 
%        ThAr          &  2010-06-15       & 05:57:02   & 580  &        5           &    9.2  &   9.2   &10.3 &  10.3   &743.32   &743.32   &  743.0   & 743.0         &4690921   &   \\      
                    &  2010-06-15       & 05:58:21   & 580    & 4000           &      \\
 %     ThAr            &  2010-06-15       & 07:06:10   & 580 &       5            &   9.1  &   9.1   &10.1   & 10.1    &743.47  & 743.47  &  742.6 &  742.6           &4690921  &   \\   
                    &  2010-06-15       & 07:07:43  & 580   & 3767           &      \\    
 %     ThAr            &  2010-06-15       & 08:11:39   & 580 &       5            &   9.0  &   9.0   &10.0   & 10.0    &743.08  & 743.08  &  742.2 &  742.2           &4690918  &   \\ 
%\hline
%       ThAr           &  2010-06-15       & 08:14:37   & 580 &       5            &   8.9 &  8.9  & 10.0  & 10.0   & 743.08   & 743.08   &  742.1 &  742.2         &4690913  &   \\   
                    &  2010-06-15       & 08:15:58   & 580 & 3767           &    \\  
%   ThAr               &  2010-06-15       & 09:19:53   & 580 &      5            &   8.9  &   8.9   & 9.9 &   9.9   & 742.98   & 742.95  &  742.2 &  742.2              &4690913  &   \\         
                    &  2010-06-15       & 09:21:40   & 580  & 3767          &       \\                  
 %   ThAr             &  2010-06-15      & 10:25:35   & 580 &       5            &  8.8  &   8.8 &  9.8 &   9.8   &743.08  & 743.08   &  742.2 &  742.2                &4690913  &   \\     
%\hline 
 %     ThAr           &  2010-06-16       & 05:25:54   & 760 &       5            &   8.2 &  8.2 & 9.2  &9.2  & 744.22   & 744.22   &  743.2         &  743.2          &  557871  &   \\   
                   &  2010-06-16       & 05:27:15  & 760  & 4020           &      \\  
 % ThAr               &  2010-06-16       & 06:35:23 & 760  &       5            &     8.1  &   8.1   & 9.0 &   9.0   & 743.88   & 743.88 &  742.9 &  742.9             & 557871  &   \\ 
                   &  2010-06-16       & 06:36:56   & 760  & 4020           &       \\  
%  ThAr               &  2010-06-16       & 07:45:04 & 760  &        5           &     7.9 &   7.9   & 8.8 &   8.9   & 743.77   & 743.77 &  742.8 &  742.8             & 557869 &   \\ 
%\hline
%  ThAr               &  2010-06-16       & 07:48:24 & 760 &         5           &    7.9  &   7.9   & 8.8 &   8.8   & 743.73   & 743.73 &  742.8 &  742.8             & 557863 &   \\  
                   &  2010-06-16       & 07:49:45  & 760 &  4000          &      \\  
                   &  2010-06-16       & 08:59:06  & 760  & 4020          &     \\  
%ThAr                 &  2010-06-16       & 10:07:14 & 760&       5            &     7.6  &   7.6  & 8.6 &   8.6   & 743.40  & 743.40 &  742.4 &  742.4             & 557858&   \\  
%\hline 

%  ThAr               &  2010-06-17      & 05:22:37 & 760 &       5            &     8.2  &   8.2 & 9.4 &   9.4   & 743.38   & 743.38 &  742.0 &  742.0            & 557871&   \\  
                   &  2010-06-17       & 05:27:54  & 760  & 4000           &  \\
                   &  2010-06-17       & 06:35:25  & 760 & 4000           &    \\
%\hline
                   &  2010-06-17       & 07:56:53  & 760  & 4000           &     \\
                   &  2010-06-17       & 09:04:47  & 760  & 4000           &      \\
% ThAr                &  2010-06-17      & 10:12:35 & 760 &      5               &     8.1 &   8.1  & 9.2 &   9.2     & 742.15    & 742.15   &  740.9   &  740.9         & 557869&   \\   
\hline
 \end{tabular}
\end{table}

\subsection{Data reduction}
We followed two independent approaches for the data reduction.
Both analysis  utilized the Common Pipeline Language version
of the various tools constituting the UVES data reduction
pipeline\footnote{See
  \url{http://www.eso.org/observing/dfo/quality/UVES/pipeline/pipe_reduc.html}
  .}. The pipeline first bias-corrects and flat-fields the quasar
exposures. The echelle orders are curved and somewhat tilted with
respect to the CCD rows/columns, so the pipeline locates them using an
exposure of a quartz lamp observed through a pinhole instead of a
standard slit. The quasar flux is then extracted with an optimal
extraction method over the several pixels in the cross dispersion
direction where the source flux is distributed. The wavelength
calibration was performed using ThAr lamp exposures; it is a
particularly important step for our purposes -- see detailed
discussion in Section \ref{sec:wavecal} below. While the CPL code
redisperses the spectra on to a linear wavelength scale by default, we
used only the original, un-redispersed flux and error array for each
echelle order in subsequent reduction steps. After wavelength
calibration, the air wavelength scale of each echelle order, of each
quasar exposure, was corrected from air to vacuum using the
\citet{Edlen:1966:71} formula, and placed in the Solar System's barycentric
reference frame using the date and time of the mid-point of the
exposure's integration.

In the first analysis approach, the extracted flux from individual
exposures was redispersed to a common linear wavelength grid and
co-added with standard routines within the ESO {\sc midas} package
using the S/N at the echelle order centres as weights. The quasar
continuum was fitted in local regions around absorption lines with
low-order polynomials. In the second analysis approach, a custom code,
\popler\footnote{Written and maintained by MTM
  \url{http://astronomy.swin.edu.au/~mmurphy/UVES_popler}} was used to
redisperse the flux from individual exposures onto a common log-linear
wavelength scale with dispersion 1.3\,\kms\,pixel$^{-1}$. The flux
from all exposures was scaled to match that of overlapping orders and
then co-added with inverse-variance weighting and cosmic ray
rejection. \popler\ was also used to automatically fit a continuum to
the final spectrum. This continuum was generally acceptable, though
some local adjustments were made using low-order polynomial fits in
the vicinity of some metal absorption lines.

\subsection{Wavelength calibration}
\label{sec:wavecal}

A variation of the $\alpha$ should manifest itself as small radial
velocity shifts between transitions. These transitions may fall
hundreds of {\AA}ngstr{\"o}ms, and many echelle order apart in the quasar
spectrum.  Since the signal for a non-zero \daa\ comes from the
relative wavelength spacing of these transitions, having an accurate
wavelength scale is crucial.  The standard method used to calibrate
the wavelength scale for a given science exposure with UVES is to take
a hollow cathode ThAr arc lamp exposure with the same echelle and
cross-disperser settings as the science exposure. Since December 2001,
ESO has implemented an automatic resetting of the Cross Disperser
encoder positions of UVES at the start of every image
acquisition\footnote{See the UVES pipeline user manual p.~78}.  This
policy was introduced to allow the use of day-time ThAr calibration
frames to save on night-time calibration overheads.  A negative aspect
of this choice is that it implies that calibrations are taken under
different conditions (e.g.~ambient temperature, pressure, telescope
pointing, telescope--instrument orientation differences etc.) from
those prevailing during the quasar observations.
%and that even a ThAr arc taken close in time to the science exposure may not fully reflect local physical changes at the telescope. 
It also implies that a ThAr exposure taken immediately before or after
a quasar exposure may not accurately reflect the wavelength scale of
the latter: the grating encoders are reset at each new acquisition
sequence, so each time the quasar or ThAr arc is acquired for a
subsequent exposure, the grating encoders, in principle, are moved.
% In the present implementation the resetting is automatically activated with the image acquisition so it is not possible to take a ThAr before the quasar avoiding the resetting %of the encoder positions. 

One of the principal aims of our new observations is to avoid possible
systematic effects introduced by this `service' calibration
plan. Therefore, ThAr exposures were taken before and after each quasar
observation during the entire observing run, though only the ThAr
frame taken after the quasar exposure was ``attached'' to the quasar
exposure: this special mode allows us to avoid the automatic resets of
grating positions between the quasar and calibration exposures.  For
this reason, the wavelength calibration of each quasar exposure
studied in this paper was performed with its corresponding
``attached'' ThAr exposure.  This procedure ensures that no mechanical
movements occur between the quasar and ThAr frames and provides the
best possible calibration for UVES. 
% A change in the encoder value of few units as often reported in the last column of the table in Appendix indicates 
% encoder resets. 

The minimum and maximum temperatures during exposures within the
spectrograph in  correspondence to  the blue arm's CCD
camera  are monitored.  During
science exposures, typical changes are $\Delta T\le0.1$\,K, with two cases
showing changes of 0.2 K. Ambient air pressure values are surveyed at
the beginning and end of the exposures and reveal pressure changes
of $\Delta P=0.2$ to 0.8 mbar.
% \footnote{Note that start and end pressure values are inverted in the UVES fits headers (TBC)}. 
The impact of the temperature and pressure variations on UVES radial
velocity stability are of 50\,\ms\ for $\Delta T = 0.3$\,K or $\Delta
P = 1$ mbar\footnote{UV-Visual Echelle Spectrograph User Manual, 2004,
  http://www.eso.org/instruments/uves/userman .}.  The measured values
assure radial velocity stability within $\approx$50\,\ms. Of course,
if such a spurious velocity shift were equally imparted to all
transitions in a given spectrum, this would not result in a spurious
shift in \daa; only velocity shifts \emph{between} transitions should
affect a measured value of \daa. According to the \citet{Edlen:1966:71} formula
for the refractive index of air, pressure and temperature changes of
0.1\,K and 1\,mbar would cause \emph{differential} velocity shifts of
$\la$15 \ms\ across the wavelength range of the spectrum studied here,
implying a negligible effect on \daa.

% However, there is not stringent correlation indicating that mechanical jumps contribute to the changes in the encoder values between different exposures. 
%These problems were not fully addressed in previous analysis performed with UVES or HIRES data for measuring fundamental constant variability. 

The use of attached ThAr lamps, together with the use of no on-chip
pixel binning, and a narrower spectrograph slit are the main
differences with respect to most of the previous observations dealing
with the search for the variability of fundamental physical constants
using UVES.

In the first analysis approach, the wavelength calibration was
established with the CPL pipeline's standard method, i.e.~extraction
of the ThAr flux along the centre of each echelle order. In the second
analysis approach, the ThAr flux was extracted using the object
profile weights from its corresponding quasar exposure. The procedure
for the selection of ThAr lines and the analysis leading to a
wavelength solution follows that outlined in
\citet{Murphy:2007:839}. Specifically, in the 390\,nm or 437\,nm frames about
400 ThAr lines were identified, and more than 55\% of them were used
to calibrate the lamp exposures.  A polynomial of the 5th order was
adopted to match the selected lines.  Residuals of the wavelength
calibrations were typically of $\approx$0.34\,\ma, or $\approx$24 \ms\
at 400\,nm and symmetrically distributed at all wavelengths.  We note
that uncertainties of the laboratory wavelengths of individual thorium
lines vary from about 15\,\ms\ up to more than 100\,\ms\ for the
poorly known lines \citep{Lovis:2007:1115,Murphy:2007:839}.
%In Fig \ref{fig:ident} and Fig \ref{fig:residuals} the line identification and the residuals for the red low CCD of the 580 nm frame are shown as an example of the wavelength calibration.

\subsection{Slit effects}
\label{sec:slit-effects}

The optical paths of the ThAr and quasar light beams through the
spectrograph are not identical and therefore there may be effects
which cannot be traced by the ThAr calibrations.  The most important
one is the difference in the slit illumination between a point like
source, such as a quasar, and the calibration lamp, which provides a
relatively uniform illumination of the slit.  Even when the point-like
source is well centered into the slit, the precise position could vary
slightly from one exposure to another, depending also on telescope
guiding and tracking. In UVES, a 1\farcs0 slit-width corresponds to
$\approx$7.5\,\kms\ when projected onto the CCD. Therefore even a
small shift of 0\farcs01 in the centroid of the light is sufficient to
induce a shift of $\sim$100\,\ms\ in the entire spectrum relative to
its corresponding ThAr calibration exposure. Exposures taken at
different times will in general show different radial velocity shifts
due to this effect. Indeed, this effect has been measured by
\citet{Wendt:2012:69} by cross-correlating 15 UVES spectra of quasar
Q0347$-$383.  These spectra showed peak-to-peak excursions of up to
$\approx$800\,\ms\ with an average deviation of 170\,\ms.

Such a `slit shift' of all metal absorption lines by the same velocity
in a single spectrum does not affect \daa. However, this effect
becomes relevant when co-adding different exposures and when comparing,
as it is for our observations, spectra obtained with different UVES
settings: the mean slit shift will vary as a function of wavelength
because of the different contributions from different settings at
different wavelengths. This can produce a systematic effect in \daa\
if absorption lines from quite different wavelength regions are
used. Also, small radial velocity shifts between exposures will
produce a slight degradation of the S/N and effective instrumental
line-shape in the final co-added spectrum.  The desire to obtain
co-added quasar spectra with high S/N precludes the use of individual
quasar exposures for analysis because they usually have too low S/N.
One exception was the very bright quasar HE\,0515$-$4414, which
provided one of the highest statistical precisions on \daa\ from an
individual absorption system so far \citep{Levshakov:2005:827}.

\begin{figure*}
\centerline{
\hbox{
  \includegraphics[width=0.45\textwidth]{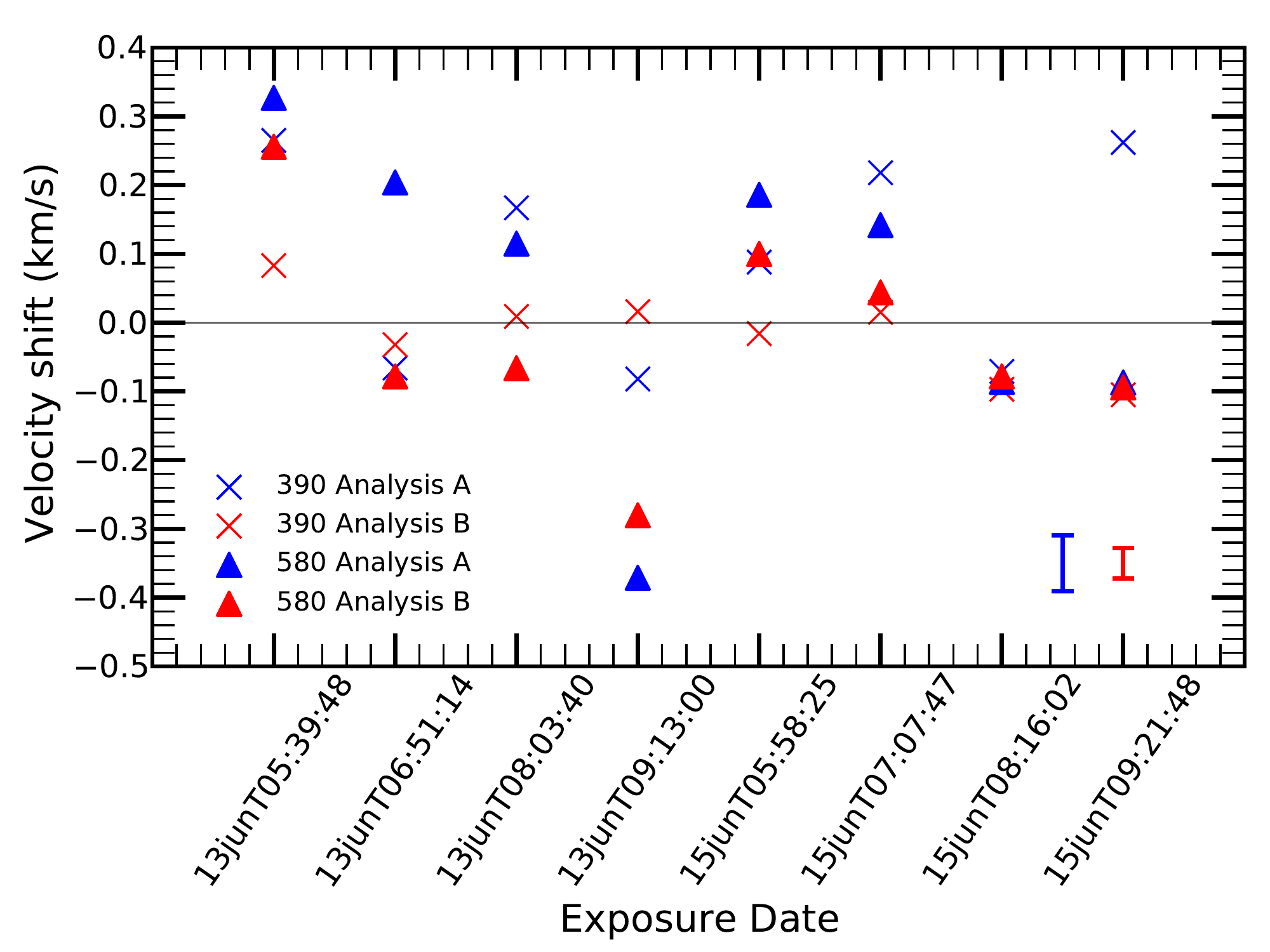}
  \hspace{0.01\textwidth}
  \includegraphics[width=0.45\textwidth]{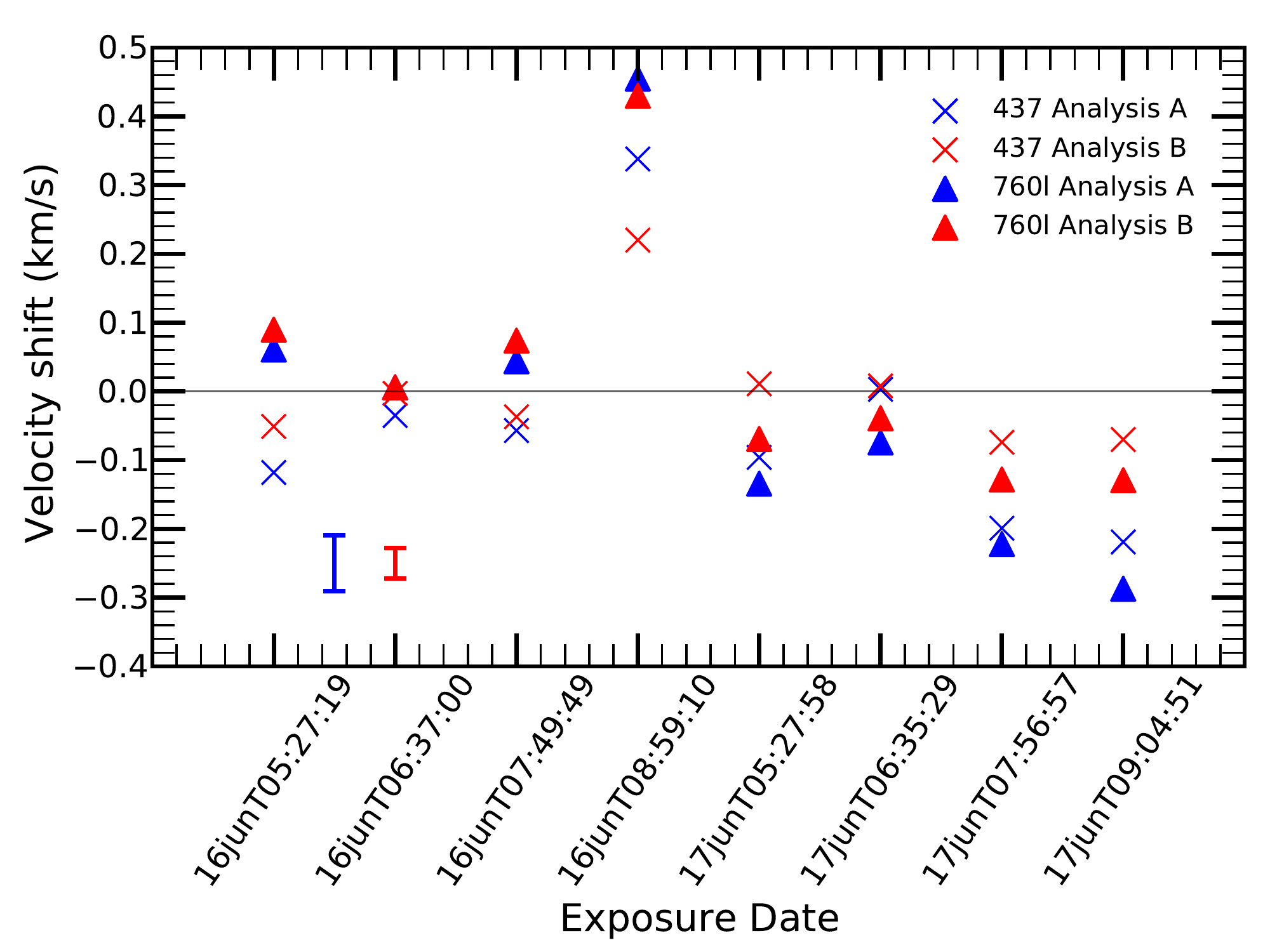}
 }
}
\caption{Radial velocity `slit shifts' between absorption lines in
  individual exposures and the same lines in the mean spectrum. Left
  panel: Shifts for the 390\,nm (crosses) and 580\,nm (triangles)
  settings.  Right panel: Shifts for the 437\,nm (crosses) and bluer
  chip of the 760\,nm (triangles) settings. In both panels, blue
  points correspond to the first analysis approach and the red points
  correspond to the second analysis approach. Representative error
  bars for each analysis approach are shown in each panel.}
\label{fig:shifts390+437ave}
\end{figure*}

In our first analysis approach we measured the relative slit shifts
of the individual quasar exposures by cross-correlating them with the
mean spectrum.
% The XCSAO code within IRAF was applied to selected wavelength intervals containing the strongest absorption features but without telluric lines or cosmic hits.
After selecting regions without telluric lines or `cosmic-ray'
features, we applied the XCSAO code from IRAF to cross-correlate the
strongest absorption features.  The two settings are summarized in
Fig.~\ref{fig:shifts390+437ave}.
% The results of the cross-correlation for the two settings are summarized in Figs \ref{fig:shifts390ave} and \ref{fig:shifts437ave}.
Despite the fact that relatively few features can be used for the
purpose, global radial velocity shifts between different exposures
were detected with high confidence.  These slit shifts are typically
of few hundred \ms\ with a maximum amplitude of $\approx$700\,\ms, in
agreement with \citet{Wendt:2012:69}.

The second analysis approach used a custom $\chi^2$ minimization
technique to compare each individual exposure with the mean spectrum
(rather than a cross-correlation method). This method is described in
detail by \citet{Evans:2013}. Its principle advantage for this
application is  that it does not require prior identification of
``strong'' absorption features, instead utilizing all absorption
features which contain enough `spectral information' to provide a
reliable shift measurement for each exposure. The slit shifts measured
via this second approach are also shown in
Fig.~\ref{fig:shifts390+437ave}. In general there is very good
agreement between the measurements from the two approaches, broadly
consistent with the typical uncertainty of $\sim$30--80\,\ms.

In Fig.~\ref{fig:shifts390+437ave} it is interesting to note that, in
general, the shifts of the two different UVES arms of a given setting
are comparable in amplitude and sign. This is expected in the case of
good optical alignment of the two separate slits of the two UVES arms
(blue and red).  This is not obvious {\it a priori} because the two
arms of the spectrograph have two different entrance slits and any
optical misalignment will produce radial velocity differences between
the two arms' spectra. \citet{Molaro:2008:559} compared the radial velocity
measured in the two arms using the sunlight reflected by an asteroid,
and came to a similar conclusion. We note that with UVES, no telescope
or instrument control is provided to control the relative slit
alignment with the desirable accuracy, and the optical alignment can,
in principle,  change from one observing run to another.

In both analysis approaches, each exposure was corrected for the slit
shifts discussed above. The 8 exposures were then combined in each
setting separately to allow overlapping regions of neighboring
settings to be compared using the same two approaches discussed above
(cross-correlation and $\chi^2$ minimization). This revealed residual
velocity shifts, referred to as `setting shifts' hereafter, between
the 390 and 437-nm co-added spectra of $-60\pm52$ and $-93\pm38$\,\ms\
in the first and second analysis approaches, respectively. The
settings shifts between the overlapping portions of the 580 and 760-nm
co-added spectra were $-7\pm48$ (first approach) and $-155\pm43$\,\ms
(second approach).
% Subsequently we performed a cross-correlation between the spectra of the two different settings by using the lines in the overlapping region to put all spectra on the same velocity scale.
Unfortunately, there are too few strong, particularly narrow
absorption features in the small overlapping wavelength range
($\sim$4800--4950\,\AA) between the 437 and 580-nm settings, so we
cannot check the setting shifts between them with adequate
precision. However, the previous results for the slit shifts between
the two UVES arms discussed above provide evidence that the shift
between the 437 and 580-nm settings should be consistent with zero
within $\sim$30\,\ms. That is, we should expect the 390-vs-437-nm and
580-vs-760-nm setting shifts to be consistent with each other within
each analysis approach, which we do observe. Indeed, for the second
analysis approach, we adopted a single setting shift of
$-124\pm41$\,\ms\ as representative of the 390-vs-437 and
580-vs-760-nm estimates. We should not expect large differences
between the setting shifts of the two different approaches, and this
also seems to be observed.

%With respect to a blind average the 390+580 nm setting has been shift by -80 $\pm$ 45, -100 $\pm$ 0.04 and -260 $\pm$ 31 for the blue and 3 red CCDs respectively and for thar 437+760 the shits are of 180 $\pm$2, 10 $\pm$ 2, and 9 $\pm$ 4 \ms for the 437 and 760 low and up respectively.
%In our cases since the average is performed on 8 spectra the final co-added spectrum is not particularly different in terms of radial velocity from that obtained from a direct average form individual source. 
% But this is only due to the fact that we have a relatively high number of exposures of the same object. The whole procedure shows how dangerous could be in relaying on few exposures which can deviate by several hundred of \ms.

In both analysis approaches, the final spectrum (in which \daa\ is to
be measured) was formed by co-adding the 8 exposures in all settings after
removing the `slit shifts' and `setting shifts' from the individual
exposures. These final spectra have a high S/N, with values up to
$\approx$100 in the continuum in most of the spectral regions of
interest. In Section \ref{ssub:systematic_error_tests} we consider the
effect of not correcting for the slit and setting shifts.

\subsection{Known systematic effects in the wavelength calibration}

The possible presence of distortions in wavelength scales of
Keck/HIRES spectra was investigated by \citet{Griest:2010:158} by comparing the
ThAr wavelength scale with that established from I2-cell observations.
In the wavelength range $\sim$5000--6200\,\AA\ covered by the iodine
cell absorptions they found a saw-tooth distortion pattern with
amplitude of typically 300 \ms\ along each echelle order.
%On top of this they found absolute offsets which can be as large as 500 -- 1000 \ms which can have a different origin as we discuss below.
The distortions are such that transitions at the order edges appear at
different velocities with respect to transitions at the order centers
when the two calibrations are compared.  \citet{Whitmore:2010:89} repeated the
same test for UVES and found similar effects, though the saw-tooth
distortion shows reduced peak-to-peak velocity variations of
$\sim$200\,\ms.  The origin of these distortion is not clear.    \citet{Wilken:2010:L16}  on spectra obtained with the   
HARPS  spectrograph at the 3.6m ESO telescope on La Silla,  and  calibrated  by means of a Laser
Frequency Comb detected  very small differences in the size of physical
pixels due to CCD manufacturing.  In general, an individual CCD sensor
is built up by means of several unit cells or blocks which are
connected to each other during the design or lay-out phase of the
development cycle of the sensor.  The block stitching process could
produce misalignments between different CCD blocks of the order of
$\sim$$10^{-2}$ of the pixel size.
The calibration polynomial cannot cope with these sudden changes in
the dispersion solution and oscillates to accommodate it.  This
``overcompensation'' will produce local calibration distortions.
Since blocks have typical sizes of 512 pixels, or $\approx$13\,\AA\ in
UVES, shifts could occur even for neighboring absorption lines, such
as the \ion{Mg}{ii} doublet, or even within a single absorption
profile where the components span several hundreds \kms\ and may fall
over two different CCD manufacturing blocks. We test for systematic
errors of this kind in Section \ref{ssub:systematic_error_tests}.

We note that \citet{Molaro:2011:167} compared solar features observed both with
HARPS and UVES and found such `intra-order distortions' in the UVES
spectrum with a typical scale of $\approx$10\,\AA.  In HARPS the
offsets were measured up to 50 \ms within one order and in UVES, where
the pixel size is a factor 3 bigger, the offsets could be a factor 3
larger.  We note also that the $\pm$40\,\ms\ inaccuracies of the HARPS
wavelength scale revealed by the Laser Frequency Comb test contribute
to the line-list constructed for use with UVES \citep{Lovis:2007:1115,Murphy:2007:839}.

\section{Many-Multiplet method and Fitting}

\subsection{Many-Multiplet method} % (fold)
\label{sub:many_multiplet_method}

The change in the rest-frame frequencies between the laboratory,
$\omega_{i}(0)$, and in an absorber at redshift $z$, $\omega_{i}(z)$,
due to a small variation in $\alpha$, i.e.~\daa $\ll 1$, is
proportional to a $q$-coefficient for that transition:
\begin{equation}\label{eq:da1}
\omega_{i}(z) \equiv \omega_{i}(0) + q_i\left[\left(\alpha_z/\alpha_0\right)^2-1\right]\,,
\end{equation}
where $\alpha_0$ and $\alpha_z$ are the laboratory and absorber values
of $\alpha$, respectively \citep{Dzuba:1999:888}.  The $q$-coefficients for the
transitions of interest are given in the last column of
Table~\ref{table:atom}.  The change in frequency is observable as a
velocity shift, $\Delta v_i$, of transition $i$.
\begin{equation}\label{eq:da2}
\hspace{1em}\frac{\Delta v_i}{c} \approx -2\frac{\Delta\alpha}{\alpha}\frac{q_i}{\omega_{i}(0)}\,.
\end{equation}
The MM method is based on the comparison of measured velocity shifts
from several transitions having different $q$-coefficients to compute
the best-fitting \daa.

When applied across a variety of different ionic species
(i.e.~different atoms or ionization stages of the same atom), the MM
method assumes that the different species traces the same material, or
at least that there are no substantial intrinsic velocity shifts
between the absorption lines of different species. Therefore, to
reduce the number of free parameters in a fit to all transitions of
all species, we impose the requirement that corresponding fitted
velocity components in different species share the same overall
redshifts (i.e.~cosmological redshift and peculiar velocity combined).
Unless there is evidence for the lines being thermally broadened, we
also assume that the lines are broadened by macro-turbulence
motions. That is, we assume that corresponding velocity components in
different species have the same fitted Doppler broadening parameter.

The laboratory atomic parameters used in our analyses -- rest-frame
frequencies, oscillatory strengths etc.~-- were taken from a new
compilation (Murphy \& Berengut, in prep.), the important details of
which are not significantly different from that used by
\citet{King:2012:3370}. These are reported in Table~\ref{table:atom}.  In
particular, the laboratory wavelength for the \ion{Fe}{ii}
$\lambda$1608 transition was taken from \citet{Nave:2011:737}. The isotopic
structures of most transitions are also reported, together with the
composite laboratory wavelength expected for a solar isotope
composition.
% In fact there is very little information about the isotopic Mg composition in these systems and when Mg is the only anchor used this makes the \daa analysis very sensitive to this assumption.

% Table atom
% \input{atom} 
 \begin{table}[ht!]
   \caption{Laboratory atomic data for the transitions used in our
     analyses, taken from Murphy \& Berengut \citep[in prep.; see also][]{King:2012:3370}.
     The third column shows the oscillator strength or, in
     the case of an isotopic or hyperfine structure component, the terrestrial isotopic
     abundance fraction or the relative hyperfine level population (in
     local thermodynamic equilibrium). An asterisk ($^*$) indicates information taken from \citet{Nave:2011:737}.}
\label{table:atom}      
\centering          
\begin{tabular}{l l r r l}     % 2 columns 
 \hline    

\multicolumn{1}{c}{Ion} & \multicolumn{1}{c}{$\lambda_0$}  & \multicolumn{1}{c}{$f$} or \%  & \multicolumn{1}{c}{$q$} &  \\
    &   \multicolumn{1}{c}{[\AA]}  &   & \multicolumn{1}{c}{[cm$^{-1}$]} & \\
\hline

\ion{Mg}{i} 2852 & 2852.962797(15) & 1.83 & 86(10)&\\
  $^{26}$\ion{Mg}{i} 2852 & 2852.959591(20)& 11.0$\%$ &  &\\
  $^{25}$\ion{Mg}{i} 2852 & 2852.961407(20) & 10.0$\%$ &  &\\
  $^{24}$\ion{Mg}{i} 2852 & 2852.963420(14) & 79.0$\%$ &  &\\
\ion{Mg}{ii} 2796 & 2796.353794(16) & 0.6155 & 211(10)&\\
 $^{26}$\ion{Mg}{ii} 2796 & 2796.3470457(4) & 11$\%$ & &\\
 $^{25}$\ion{Mg}{ii} 2796 & 2796.348269(50) & 4.2$\%$ & &\\
 $^{25}$\ion{Mg}{ii} 2796 & 2796.352784(50) & 5.8$\%$ & &\\
 $^{24}$\ion{Mg}{ii} 2796 & 2796.3550990(4) & 79.0$\%$ & &\\
\ion{Mg}{ii} 2803 & 2803.530983(16) & 0.3058 & 120(2)&\\
  $^{26}$\ion{Mg}{ii} 2803 & 2803.5242094(4) & 11$\%$ &  &\\
  $^{25}$\ion{Mg}{ii} 2803 & 2803.525314(50) & 4.2$\%$ &  &\\
  $^{25}$\ion{Mg}{ii} 2803 & 2803.530004(50) & 5.8$\%$ &  &\\
  $^{24}$\ion{Mg}{ii} 2803 & 2803.5322972(4) & 79.0$\%$ &  &\\
\ion{Al}{ii}  1670 & 1670.78861(11) & 1.74 & 270(30)&\\
\ion{Al}{iii}  1854  &  1854.717941(34)& 0.559 & 464(30)&\\
          ~~~~~$F$=2       &  1854.708966(28)&  41.7& &\\
          ~~~~~$F$=3       &  1854.724704(21)&58.3  & &\\
\ion{Al}{iii}  1862  &  1862.791127(69)& 0.278 & 216(30)&\\
          ~~~~~$F$=2       &  1862.780325(52)&  41.7& &\\
          ~~~~~$F$=3      &  1862.798581(42)&58.3  & &\\
\ion{Si}{ii} 1526 & 1526.706980(16) & 0.133 & 50(30) &\\
$^{30}$\ion{Si}{ii} 1526   & 1526.7072550   & 3.1$\%$ & &\\
$^{29}$\ion{Si}{ii} 1526   & 1526.7071150   & 4.7$\%$ & &\\
$^{28}$\ion{Si}{ii} 1526   & 1526.7069637   & 92.2$\%$& &\\
\ion{Fe}{ii} 1608         & 1608.45081(7)$^*$ & 0.0577 &$-$1299(300) &\\
$^{58}$\ion{Fe}{ii} 1608  & 1608.4519539    & 0.3$\%$ & &\\
$^{57}$\ion{Fe}{ii} 1608  & 1608.4514182    & 2.1$\%$ & &\\
$^{56}$\ion{Fe}{ii} 1608  & 1608.4508635    & 91.8$\%$& &\\ 
$^{54}$\ion{Fe}{ii} 1608  & 1608.4496923    & 5.8$\%$ & &\\
\ion{Fe}{ii} 2344         & 2344.212747(76) & 0.114  &1210(150)& \\
$^{58}$\ion{Fe}{ii} 2344  & 2344.2113616     & 0.3$\%$ & &\\   
$^{57}$\ion{Fe}{ii} 2344  & 2344.2120103     & 2.1$\%$ & &\\
$^{56}$\ion{Fe}{ii} 2344  & 2344.2126822     & 91.8$\%$& &\\ 
$^{54}$\ion{Fe}{ii} 2344  & 2344.2141007     & 5.8$\%$ & &\\
\ion{Fe}{ii} 2374         & 2374.460064(78)    &0.0313 & 1590(150)&\\
$^{58}$\ion{Fe}{ii} 2374  & 2374.4582998    & 0.3$\%$ & &\\
$^{57}$\ion{Fe}{ii} 2374  & 2374.4591258    & 2.1$\%$ & &\\
$^{56}$\ion{Fe}{ii} 2374  & 2374.4599813    & 91.8$\%$& &\\ 
$^{54}$\ion{Fe}{ii} 2374  & 2374.4617873    & 5.8$\%$ & &\\
\ion{Fe}{ii} 2382         & 2382.763995(80) & 0.320& 1460(150)&\\
$^{58}$\ion{Fe}{ii} 2382  & 2382.7622294    & 0.3$\%$ & &\\
$^{57}$\ion{Fe}{ii} 2382  & 2382.7630560    & 2.1$\%$ & &\\
$^{56}$\ion{Fe}{ii} 2382  & 2382.7639122    & 91.8$\%$& &\\ 
$^{54}$\ion{Fe}{ii} 2382  & 2382.7657196    & 5.8$\%$ & &\\
\ion{Fe}{ii} 2586         & 2586.649312(87)  &0.0691& 1490(150) &\\
$^{58}$\ion{Fe}{ii} 2586  & 2586.6475648    & 0.3$\%$ & &\\
$^{57}$\ion{Fe}{ii} 2586  & 2586.6483830    & 2.1$\%$ & &\\
$^{56}$\ion{Fe}{ii} 2586  & 2586.6492304    & 91.8$\%$& &\\ 
$^{54}$\ion{Fe}{ii} 2586  & 2586.6510194    & 5.8$\%$ & &\\
\ion{Fe}{ii} 2600         & 2600.172114(88)& 0.239& 1330(150) & \\  
$^{58}$\ion{Fe}{ii} 2600  & 2600.1703603    & 0.3$\%$ & &\\   
$^{57}$\ion{Fe}{ii} 2600  & 2600.1711816    & 2.1$\%$ & &\\
$^{56}$\ion{Fe}{ii} 2600  & 2600.1720322    & 91.8$\%$& &\\ 
$^{54}$\ion{Fe}{ii} 2600  & 2600.1738281    & 5.8$\%$ & &\\
\hline    
\end{tabular}
\end{table}

\subsection{Selection of transitions} % (fold)
\label{sub:selection_of_transitions}

To build the best model of the velocity structure of an absorption
system, we try to include as many different transitions as possible.
However, the presence of contaminating absorption lines could strongly
skew the analysis, so we did not include transitions that were blended
with absorption features of other systems at other redshifts.  For the
two different analysis approaches we took two approaches to deal with
sky (telluric) lines.  The first approach was to remove the regions
affected by the telluric lines if their presence could be established
based on the spectrum itself, or through comparison with a standard
star spectrum taken during our observations. The second approach was
more conservative, completely rejecting transitions which had any
possible telluric line contamination, even if only in a small part of
the velocity structure of the transition. Finally, in neither approach
did we consider any transitions which fall in the Ly$\alpha$ forest
portion of the spectrum for the analysis of \daa, even if part of
their velocity structure presented no evidence of contamination by
broad Ly$\alpha$ lines.

\subsection{Absorption-line fitting} % (fold)
\label{sub:fitting}

The model fitting of the absorption profile is performed by means of the non-linear least squares Voigt profile fitting program VPFIT by Bob Carswell and John Webb \footnote{The 9.5 version of the VPFIT manual is available at \url{ftp://ftp.ast.cam.ac.uk/pub/rfc/vpfit9.5.pdf}}.  This code has been widely used for the absorption line analysis and has been modified to incorporate $\alpha$ as a free parameter in the fit \citep{Murphy:2003:609}. For a given transition to be fitted, the model is specified by defining the number of components, the spectral region on which the fitting is performed and the instrumental profile.  The analysis in both approaches proceeds in two basic steps.  The first one consists in obtaining a fiducial model for the absorption while keeping \daa\ fixed at zero. That is, the number of velocity components and their distribution across the absorption complex, is refined without any influence on or from a varying \daa\ parameter. Once the model for the absorption is derived, giving a satisfactory fit of the data with all residuals consistent with the error arrays, \daa\ is introduced as a new free parameter to provide a better fit to the data.

The statistical uncertainty in \daa\ is determined from the relevant diagonal term of the covariance matrix of the best-fitting solution. This is multiplied by $\sqrt{\chi^2_\nu}$, where $\chi^2_\nu$ is the $\chi^2$ between the model and data per degree of freedom, $\nu$, to yield more reliable uncertainties \citep{Press:1992:1}. The accuracy and reliability of these statistical uncertainties has been confirmed with detailed Monte Carlo and also Monte Carlo Markov Chain simulations  \citep{King:2009:864,Murphy:2002:1}. This approach is successful because the covariance between the \daa\ parameter and other parameters in the fit is minimal. However, this is not always true for parameters of individual components in the fit. In particular, the column density and Doppler broadening parameters of an individual component are covariant, and strongly covariant with those parameters of close-blended neighboring components. In the final absorption profile models presented in Tables 4--8, and described in the following sections, we present only statistical uncertainties derived from the diagonal terms of the covariance matrix for parameters of individual components.

Finally, during the fitting process, we set a lower limit on the Doppler parameter for individual components of $b=0.5$\,\kms, which is well below the FWHM resolution of the spectra ($\sim$5.5\,\kms). These unresolved components generally have formal uncertainties (quoted in Tables 4--8) exceeding $0.5$\,\kms. In these cases, and for other components with $b$-parameters uncertainties exceeding their best-fitted value, it should be recognized that, while $b$ is formally consistent with zero, this does not indicate that the component is redundant in the fit.

\subsection{Instrument profile} % (fold)
\label{sub:instrument_profile}

The full-width-at-half-maximum (FWHM) of the instrumental profile of
UVES varies by $\approx$20\% along every echelle order due to the
spectrograph anamorphosis.
%One example for the 580 nm low CCD is shown in Fig \ref{fig:fwhm}. 
Therefore, the FWHM was computed by averaging the width of unblended
thorium emission lines in the wavelength region around the metal
absorption lines used to derive \daa.
% As a result, the FWHM is taken differently from one line to the other.
That is, the FWHM was calculated separately for each transition.
Moreover, it is known that the FWHM of ThAr lines produced by a
fully-illuminated slit is slightly broader than that produced by a
point source which under-fills the slit. Simple models suggest that,
even with FWHM seeing equivalent to the slit-width, the reduction in
the FWHM resolution is $\approx$10\%.  The values used in the present
analysis are those given in Table~\ref{table:fwhm} reduced by 10\%.

% [The independent code used by Levshakov et al has been also used for comparison. This provides the radial velocity difference form the different lines and therefore the \daa is an external parameter in the fitting procedure.]

% Table fwhm
% \input{fwhm}
\begin{table*}
  \caption{Local FWHM of the instrumental profile and local S/N measured
    in the continuum around each transition for all 5 absorbers studied here.}
\label{table:fwhm}      
\centering          
\begin{tabular}{l r  c r c r c  r c  r c }     %  columns 
 \hline    
 & \multicolumn{2}{c}{\zabs=1.6919} &  \multicolumn{2}{c}{\zabs=1.6279}  & \multicolumn{2}{c}{\zabs=1.5558}  & \multicolumn{2}{c}{\zabs=0.94} & \multicolumn{2}{c}{\zabs=0.7866}\\
 \hline  
Ion &  S/N  &   FWHM & S/N  & FWHM &   S/N &  FWHM  & S/N& FWHM & S/N & FWHM   \\     
    &            &  [\kms]   &            &  [\kms]                                 &          & [\kms]     &        & [\kms]	 &    &  [\kms]   \\
\hline
\ion{Fe}{ii} 1608 & 98 & 5.22 &    &   && &&\\
\ion{Fe}{ii} 2344  &117 & 5.51 & 110 & 5.99 &    121 & 5.61&& \\
\ion{Fe}{ii} 2374 & 103&  5.82   && && && \\
\ion{Fe}{ii} 2383  &130 &5.75 & 113 & 5.70 &    124& 5.81  &88 & 5.38 &  106 &  5.21    \\  
\ion{Fe}{ii} 2600  & 88 & 5.43& 84 &  5.40    & 84 & 5.87  & 71 &6.06      & 73   & 5.38 \\ 
 
\ion{Si}{ii} 1526 &152 &5.24  &85 &  5.35   & 60 & 5.35 &&   \\
 
Al II 1670 &77 & 5.50& 78 & 5.21   & 100 & 5.34 && \\

\ion{Mg}{ii} 2796 &78 & 5.22&  96 & 5.42  & 97 &  5.42  &53 & 5.97 & 65 & 5.58 \\		
\ion{Mg}{ii} 2803 && & 73  &5.42  & 77 & 5.42 & 51 & 5.95                 & 65 & 5.58             \\  
 
  \hline

\end{tabular}
\end{table*}  

%%%%%%%%%%%%%%%%%%%%%%%%%%%%%%%%%%%%%%%%%%%%%%%%%%%%%%%%%

% \section{System Analysis}
\section{Analysis of absorption systems}

We present below the analysis of the 5 absorption systems towards
HE\,2217$-$2818 in which we can measure \daa. We focus first on the
highest-redshift absorber, that at $\zabs=\systemARedshift$, because this
provides by-far the highest statistical precision on \daa\ and,
therefore, requires the most comprehensive analysis and consideration
of underlying systematic errors. We therefore describe in detail our
two different analysis approaches for this system. For the other
absorbers, we do not present all details of systematic error checks
because the statistical uncertainly dominates the total error budget;
we present only the results one analysis approach (the second) for
simplicity.

\subsection{Absorption system at $\zabs=\systemARedshift$}

\begin{figure*}
\begin{center}
\includegraphics[width=0.75\textwidth]{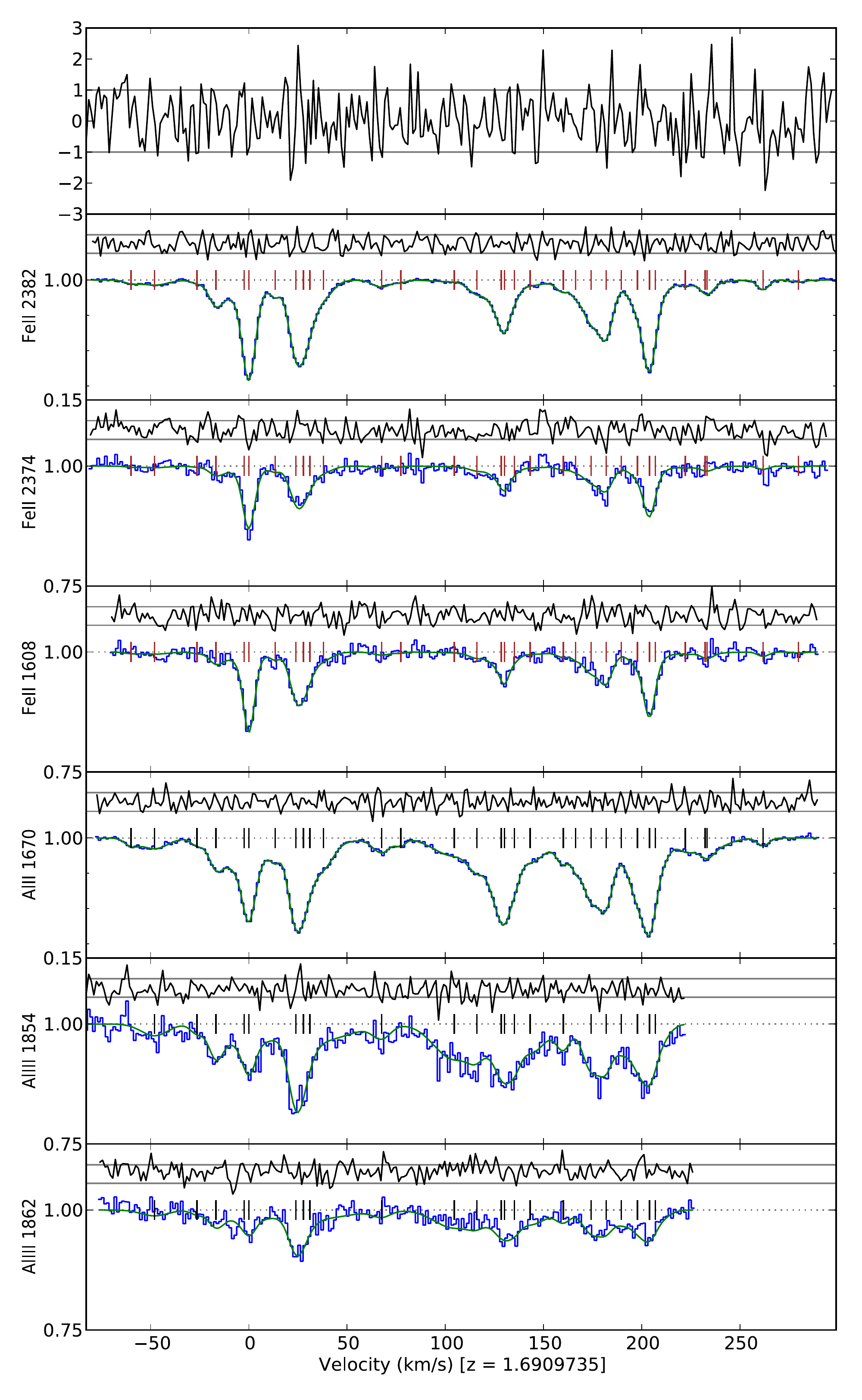}
\caption{Transitions in absorption system at $\zabs=\systemARedshift$ used to
  derive \daa\ in our second analysis approach. The Voigt profile
  model (green line) is plotted over the data (blue histogram). The
  velocity of each fitted component is marked with a vertical line and
  the residuals between the data and model, normalised by the error
  spectrum, are shown above each transition. The top panel shows the
  composite residual spectrum -- the mean spectrum of the normalised
  residuals for all transitions shown -- in units of $\sigma$.}
\label{fig:sys169}
\end{center}
\end{figure*}

The system at redshift $\zabs=\systemARedshift$ shows a quite complex velocity
structure with many separate strong, narrow features spread over
$\sim$250\,\kms. The transitions observed red-wards of the Ly$\alpha$
forest in our spectral range, and which are potentially useful for
measuring \daa\ are \ion{Al}{ii} $\lambda\lambda$1670, \ion{Al}{ii}
$\lambda\lambda$1854, 1862, \ion{Mg}{ii} $\lambda\lambda$2796 and
\ion{Fe}{ii} $\lambda\lambda$1608, 2344, 2374, 2382, 2586,
2600. Unfortunately, the detected \ion{Mg}{ii} $\lambda$2796 transition
cannot be used because partially contaminated by a strong sky emission
at $\approx$7526.8\AA. The \ion{Mg}{ii} $\lambda$2803 and \ion{Mg}{i}
$\lambda$2852 transitions were not detected because they fall into the
gap between the two red-arm CCDs in the 760-nm setting. The subset of
these transitions used to constrain \daa\ in the second analysis
approach is shown in Fig.~\ref{fig:sys169}.

Figure \ref{fig:sys169} shows that the velocity structure and relative
strength of the different features is very similar in all the
low-ionization transitions. The \ion{Al}{iii}
$\lambda\lambda$1854/1862 doublet is not observed with high S/N and is
not particularly strong. As a higher ionization species than the other
transitions, including it in the analysis may introduce a stronger
sensitivity to possible ionization differences between
species. However, no evidence for differences in velocity structure
are evident in Fig.~\ref{fig:sys169}, even though the relative
strength of some features differs from the lower-ionization
transitions. In general, some features are quite weak even in the
stronger transitions and so they are not detectable in the weaker
transitions. Note that, even by cursory visual inspection, it is clear
that the stronger features are not symmetric and particularly narrow
(unresolved); instead, their breadth and asymmetries reveal unresolved
complex velocity structure.

% Table sys169
% \input{newsys169}
\begin{table*}[h!]
  \caption{Multi-component Voigt profile model and $\chi^2$ minimization results for the absorber at $\zabs=\systemARedshift$. The redshift ($z$), Doppler broadening parameter ($b$) and relevant column densities for each velocity component are presented, along with their 1-$\sigma$ uncertainties from {\sc vpfit} are provided in the main section of the table. The fitted value for \daa, its 1-$\sigma$ statistical uncertainty (see text for systematic error component) and the final $\chi^2$ per degree of freedom, $\chi^2_\nu$, are provided in the lower part.}
\label{table:sys169}      
% \centering          
\begin{center}
\begin{tabular}{ cccccc cc cc}     % 5 columns 
%\begin{tabular}{lrrrrrr rrrr }     % 12 columns 

\hline  
 $z$  &  $\sigma_z$ & $b$  & $\sigma_{b}$   &  \LogN{Fe}{ii} &  $\sigma_N$   & \LogN{Al}{ii} &  $\sigma_N$ & \LogN{Al}{iii} & $\sigma_N$ \\
      &  & [\kms]     & [\kms] &    [\cm]  & [\cm] & [\cm] & [\cm] & [\cm] & [\cm] \\          
 \hline      
 1.6904350  &  0.0000070  &  3.7  &  1.4  &  10.99  &  0.18          &         &           &         &      \\ 
 1.6905417  &  0.0000087  &  9.0  &  1.4  &  11.50  &  0.07    &  11.24  &  0.06     &  11.19  &  0.10      \\ 
 1.6907366  &  0.0000068  &  3.1  &  1.4  &  10.92  &  0.16    &  10.72  &  0.11     &  10.77  &  0.20      \\ 
 1.6908227  &  0.0000019  &  3.8  &  0.6  &  11.87  &  0.08    &  11.33  &  0.12     &  11.33  &  0.09      \\ 
 1.6909516  &  0.0000071  &  11.7  &  1.9 &  12.27  &  0.04   &  11.93  &  0.04     &  11.56  &  0.13      \\ 
 1.6909735  &  0.0000005  &  2.6  &  0.1  &  12.67  &  0.01    &  11.78  &  0.03     &  11.21  &  0.11      \\ 
 1.6910942  &  0.0000022  &  1.2 &  1.1  &  11.46  &  0.07    &  10.93  &  0.09           &         &      \\ 
 1.6911884  &  0.0000037  &  4.1  &  0.6  &  12.34  &  0.44    &  11.92  &  0.28     &  11.62  &  0.22      \\ 
 1.6912232  &  0.0000218  &  5.6  &  1.3  &  12.51  &  0.31    &  11.84  &  0.35     &  11.38  &  0.39      \\ 
 1.6912519  &  0.0000317  &  25.5  &  6.5 &  10.90  &  0.99   &  11.45  &  0.14     &  11.82  &  0.08      \\ 
 1.6913146  &  0.0000079  &  6.0  &  0.9  &  11.87  &  0.11    &  11.46  &  0.10           &         &      \\ 
 1.6915804  &  0.0000036  &  5.3  &  0.8  &  11.45  &  0.05    &  11.17  &  0.05     &  11.04  &  0.15      \\ 
 1.6916675  &  0.0000059  &  1.7  &  1.8  &  10.75  &  0.18    &  10.60  &  0.13           &         &      \\ 
 1.6919120  &  0.0000159  &  13.8  &  1.8  &  11.18  &  0.15   &  11.65  &  0.07     &  11.85  &  0.07      \\ 
 1.6920159  &  0.0000080  &  5.2  &  1.0  &  11.71  &  0.11    &  11.45  &  0.15     &  11.24  &  0.22      \\ 
 1.6921268  &  0.0000167  &  6.0  &  1.8  &  12.27  &  0.20    &  12.08  &  0.20     &  11.63  &  0.23      \\ 
 1.6921421  &  0.0000025  &  0.5  &  0.3  &  11.83  &  0.14    &  11.24  &  0.39     &  10.50  &  0.68      \\ 
 1.6921871  &  0.0000155  &  2.4  &  3.4  &  11.27  &  1.24    &  11.11  &  1.10     &  11.05  &  0.55      \\ 
 1.6922586  &  0.0000231  &  10.4  &  3.4  &  11.59  &  0.19   &  11.67  &  0.16     &  11.64  &  0.16      \\ 
 1.6924106  &  0.0000037  &  3.7  &  0.8  &  11.57  &  0.07    &  11.33  &  0.09     &  11.22  &  0.10      \\ 
 1.6924667  &  0.0000040  &  0.5  &  2.0  &  11.30  &  0.17    &  11.14  &  0.58           &         &      \\ 
 1.6925374  &  0.0000061  &  5.3  &  1.1  &  12.29  &  0.09    &  11.92  &  0.09     &  11.61  &  0.09      \\ 
 1.6926068  &  0.0000029  &  3.6  &  0.5  &  12.31  &  0.08    &  11.82  &  0.11     &  11.43  &  0.14      \\ 
 1.6926757  &  0.0000106  &  3.4  &  2.5  &  11.12  &  0.52    &  11.07  &  0.36     &  11.19  &  0.22      \\ 
 1.6927497  &  0.0000095  &  4.7  &  1.1  &  12.04  &  0.30    &  11.84  &  0.25     &  11.50  &  0.32      \\ 
 1.6928042  &  0.0000010  &  2.6  &  0.3  &  12.54  &  0.10    &  11.89  &  0.18     &  11.22  &  0.46      \\ 
 1.6928307  &  0.0000648  &  6.7  &  4.8  &  12.01  &  0.65    &  11.69  &  0.62     &  11.44  &  0.60      \\ 
 1.6929678  &  0.0000116  &  4.0  &  3.3  &  11.09  &  0.37    &  10.76  &  0.53           &         &      \\ 
 1.6930588  &  0.0000248  &  12.3  &  2.4  &  11.45  &  0.22   &  11.47  &  0.13           &         &      \\ 
 1.6930679  &  0.0000026  &  3.1  &  0.9  &  11.52  &  0.10    &  10.77  &  0.22           &         &      \\ 
 1.6933236  &  0.0000021  &  2.7  &  0.5  &  11.42  &  0.03    &  10.73  &  0.06           &         &      \\ 
 1.6934849  &  0.0000126  &  4.0  &  2.5  &  10.82  &  0.15          &         &           &         &      \\ 

\hline
\daa\ & $\sigma_{\rm stat}$ & $\chi^2_\nu$ &   &&&&   \\
{[ppm]} & {[ppm]}&              &   &&&&  \\
\hline
\wsystemAdaaTwoPlaces & \wsystemAdaaStatisticalErrorTwoPlaces & \wsystemAdaaChiSquare &&&& \\  % sysa = 1.69
\hline
 \end{tabular}

\end{center}
\end{table*}
% \JBcom{Paolo: you mentioned having your model included.}

% Highly ionized species as \ion{C}{iv} $\lambda$ are also present and particularly strong.
% They fall at a rather central velocity around -150 \kms where the low ionization gas is almost absent and shows a highly saturated profile.
% The \ion{Si}{ii} lines are either strongly saturated, such as \ion{Si}{ii} $\lambda$1260, or contaminated by Ly$\alpha$ interlopers of the Lyman forest such as the \ion{Si}{ii} $\lambda$1526, and are excluded from the \daa analysis. 
% In the \ion{Si}{ii} $\lambda$1526 there is a clear Ly $\alpha$ interloper at about V$\approx$-100 \kms.
% Small absorptions could evade detection and the line has not been considered completely safe for the analysis.

Careful inspection of the spectrum of a fast rotating star allowed us
to precisely identify the position of telluric features. The
\ion{Fe}{ii} $\lambda$2344, \ion{Fe}{ii} $\lambda$2586 and
\ion{Fe}{ii} $\lambda$2600 all fall in a spectral region affected by
telluric lines. Removing telluric absorption features is quite a
difficult task because they vary with the physical conditions of the
Earth's atmosphere, thereby introducing time and line-of-sight
dependencies. We did not attempt a correction but instead we adopted
two different approaches to deal with this.

In the first analysis approach we excluded the spectral portions
around the telluric lines in these transitions but we included the
remaining portions in the \daa\ analysis. For \ion{Fe}{ii}
$\lambda$2344 we considered only the intervals 6310.5--6311.5 and
6312.25--6312.7\,\AA. For \ion{Fe}{ii} $\lambda$2586 we considered
only the intervals 6959.6--6961.5 and 6964.4--6965.75\,\AA\ and for
\ion{Fe}{ii} $\lambda$2600 the intervals 6996.7--7000.5 and
7001.55--7002.79\,\AA. Our second analysis approach only considers the
transitions which are not affected by telluric or any other
contaminating lines. These are the three \ion{Fe}{ii} transitions
$\lambda\lambda$1608, 2374, 2382, the \ion{Al}{ii} $\lambda$1670
transition and the \ion{Al}{III} $\lambda\lambda$1854/1862 doublet, as
shown in Fig.~\ref{fig:sys169}.

The models of the absorption profiles were built up, step-by-step, by
adding components at different velocities in all fitted species
simultaneously and minimizing $\chi^2$ at each step with \daa\ fixed
at zero. The absorber's velocity structure is quite complex and our
first and second analysis approaches used 30 and 32 velocity
components, respectively, to obtain statistically acceptable fits with
$\chi^2$ per degree of freedom, $\chi^2_\nu\approx1$. This large
number of velocity components is consistent the aforementioned
characteristics of the profile: the five main features all show
asymmetries and require at least 2 or 3 components each to adequately
fit them. Additional components are needed to account for weak
absorption between these main features; the need for these components
is clear from inspection of the line profiles of the optically thick
transitions. We note that these weak components do not influence the
value of \daa\ significantly. The first analysis approach excluded two
of these components, i.e.~the first approach represents a `minimally
complex' model. Conversely, the second approach required that the
chosen model was that which minimized $\chi^2_\nu$. The 32-component
model is provided in Table \ref{table:sys169}, with the final
$\chi^2_\nu=1.20$.

After the number and nominal relative velocities of the fitted
components are finalised, \daa\ is introduced as an additional free
parameter and determined, along with all other free parameters, in the
$\chi^2$ process. The two analysis approaches give the following
best-fit values and 1-$\sigma$ statistical uncertainties for \daa:
\begin{equation}\label{eq:res169}
  \daa=\left\{
  \begin{array}{c l}
    \ReportStatisticalError{\psystemAdaaOnePlace}
      {\psystemAdaaStatisticalErrorOnePlace}\,{\rm ppm} 
      & \mbox{(first approach)},\\
    \ReportStatisticalError{\wsystemAdaaOnePlace}
      {\wsystemAdaaStatisticalErrorOnePlace}\,{\rm ppm} 
      & \mbox{(second approach).}
  \end{array}
  \right.
\end{equation}
The main constraints on \daa\ are contributed by the combination of
the high S/N transitions with diverse $q$-coefficients, i.e.~those of
\ion{Al}{ii} and \ion{Fe}{ii}, particularly the \ion{Fe}{ii}
$\lambda$1608 transition which has a $q$-coefficient with opposite
sign to the others. The results from the two analysis approaches in
equation (\ref{eq:res169}) appear at first to be rather
inconsistent. This may indicate that important systematic errors are
present in the calibration and analysis steps, since these constitute
the main difference between the two approaches. To explore possible
systematic errors in more details in Section
\ref{ssub:systematic_error_tests} below.

One notable difference between the two analysis approaches is that the
first includes 6 \ion{Fe}{ii} transitions, or parts thereof (after
masking of telluric features), thereby making an analysis using just
those transitions feasible. Using just one ionic species alleviates
the possibility that ionization effects may systematically affect
\daa\ \citep[e.g.][]{Levshakov:2005:827}.  Of
course, in the case of \ion{Fe}{ii}, the precision and accuracy of the
result depend heavily on the \ion{Fe}{ii} $\lambda$1608 transition and
any possible systematic errors applying to the data or wavelength
calibration relevant to it. To conduct this Fe-only test, we
restricted the region analyzed in each transition to the velocity
range $-35$ to $+250$\,\kms\ in Fig.~\ref{fig:sys169}. This
effectively removed the 2 velocity components in the model for the
first approach corresponding to the weak features at the blue and red
flanks of the absorber. A further 2 velocity components from that
model were discarded as unnecessary by {\sc vpfit} because the high
S/N \ion{Al}{ii} $\lambda$1670 transition had been removed. This left
the model with 26 components. Furthermore, we derived \daa\ on an
different  reduction of the spectrum, with a slightly different binning
and exposure combination approach to those presented above. The result
based on this Fe-only analysis is $\daa = 
  \ReportStatisticalError{\psystemAsidamdaaTwoPlaces}
  {\psystemAsidamdaaStatisticalErrorTwoPlaces}$\,ppm. The
difference between this result and that from the first approach in
equation (\ref{eq:res169}) demonstrates that systematic effects may
play an important role in the total uncertainty in \daa. We turn to
quantifying the systematic error budget for this absorber below.

\subsubsection{Systematic error tests} % (fold)
\label{ssub:systematic_error_tests}

We ran a number of tests on the second analysis approach to uncover
the sensitivity of the result to various systematic effects. For each
of these tests, we find the maximum deviation between the second
approach's \daa\ value in equation (\ref{eq:res169}) and that given
by the test, adding the results in quadrature to estimate the total
systematic error budget.

Firstly, to test the convergence of the fitting procedure, we
restarted the $\chi^2$ minimization from its final solution but with
starting values of \daa\ set to $+1\sigma$ and $-1\sigma$ away from
its fiducial value. Both these test fits converged to \daa\ values
very close to that in equation (\ref{eq:res169}). The largest
difference was 0.03\,ppm, which we added in quadrature to the total
systematic error budget.

The second test investigated the effect that co-adding the individual
exposures into a final spectrum has on the absorption line centroids
and, therefore, \daa.  The software used to combine the spectra in the
second analysis approach, \popler, first redisperses all echelle
orders, from all exposures, onto a common wavelength scale with a
constant pixel size in velocity space. A dispersion of
1.30\,\kms\,\pixel\ was used for the final spectrum. To test the
effect of redispersing and co-adding the individual exposures, we
produced four new co-added spectra with slightly different dispersion
values -- 1.28, 1.29, 1.31 and 1.32\,\kms\,\pixel\ -- which ensures
that the pixel grids will be very different for each transition from
spectrum to spectrum. The values of \daa\ for these four different
spectra varied by at most 0.54\,ppm, and this was added in quadrature
to the total systematic error budget.

The third test quantified the possible effect the `intra-order'
distortions of the UVES wavelength scale found by \citet{Whitmore:2010:89} may
have on \daa. For each echelle order, of each exposure, we applied a
saw-tooth distortion to the existing wavelength scale and then co-added
them in the usual way. The distortion applied had a peak-to-peak
amplitude of 200\,\ms, varying from $-100$\,\ms\ at the order edges to
$+100$\,\ms\ and the order centres. We also produced a second
distorted spectrum where this saw-tooth pattern was reversed. Of
course, we cannot be sure that the effect identified by 
\citet{Whitmore:2010:89}
is present in our spectra, or that it has the same pattern or
magnitude in our spectra, but the two distorted spectra produced in
this way allow us to gauge the size of the effect on \daa. Refitting
for \daa\ on these distorted spectra gives a maximum deviation between
the distorted and fiducial results of 0.56\,ppm.

The final test was aimed at understanding the effect that the `slit
shifts' and `setting shifts' discussed in Sec.~\ref{sec:slit-effects}
have on \daa.  The most straightforward test is to determine \daa\
using spectra in which these corrections have \emph{not} been applied
in the second analysis approach. Unsurprisingly, not applying the
setting shift ($-124\pm41$\,\ms) produces largest difference in \daa,
1.2\,ppm, thus demonstrating that, were these effects not corrected,
they would have caused a significant systematic error, about 50\% of
the statistical error. However, while we have corrected for these
effects, we can not have done so perfectly, and errors in our
corrections will propagate as systematic errors in \daa. The relative
uncertainty on the setting shift is 33\%, so we expect that the
residual systematic error in this case is of order 0.4\,ppm.

After adding the above estimates of the systematic error in
quadrature, the total systematic error budget is 
\wsystemAdaaSystematicErrorTwoPlaces\ ppm. 
That is, the final result from the second analysis approach is
\begin{equation}
\daa = \,\ReportStatisticalandSystematicError{\wsystemAdaaOnePlace}
  {\wsystemAdaaStatisticalErrorOnePlace}
  {\wsystemAdaaSystematicErrorOnePlace}\mbox{\,ppm}\,.
\end{equation}

\subsection{Absorption system at $\zabs=\systemBRedshift $}

\begin{figure}
  \centerline{\includegraphics[width=\columnwidth]{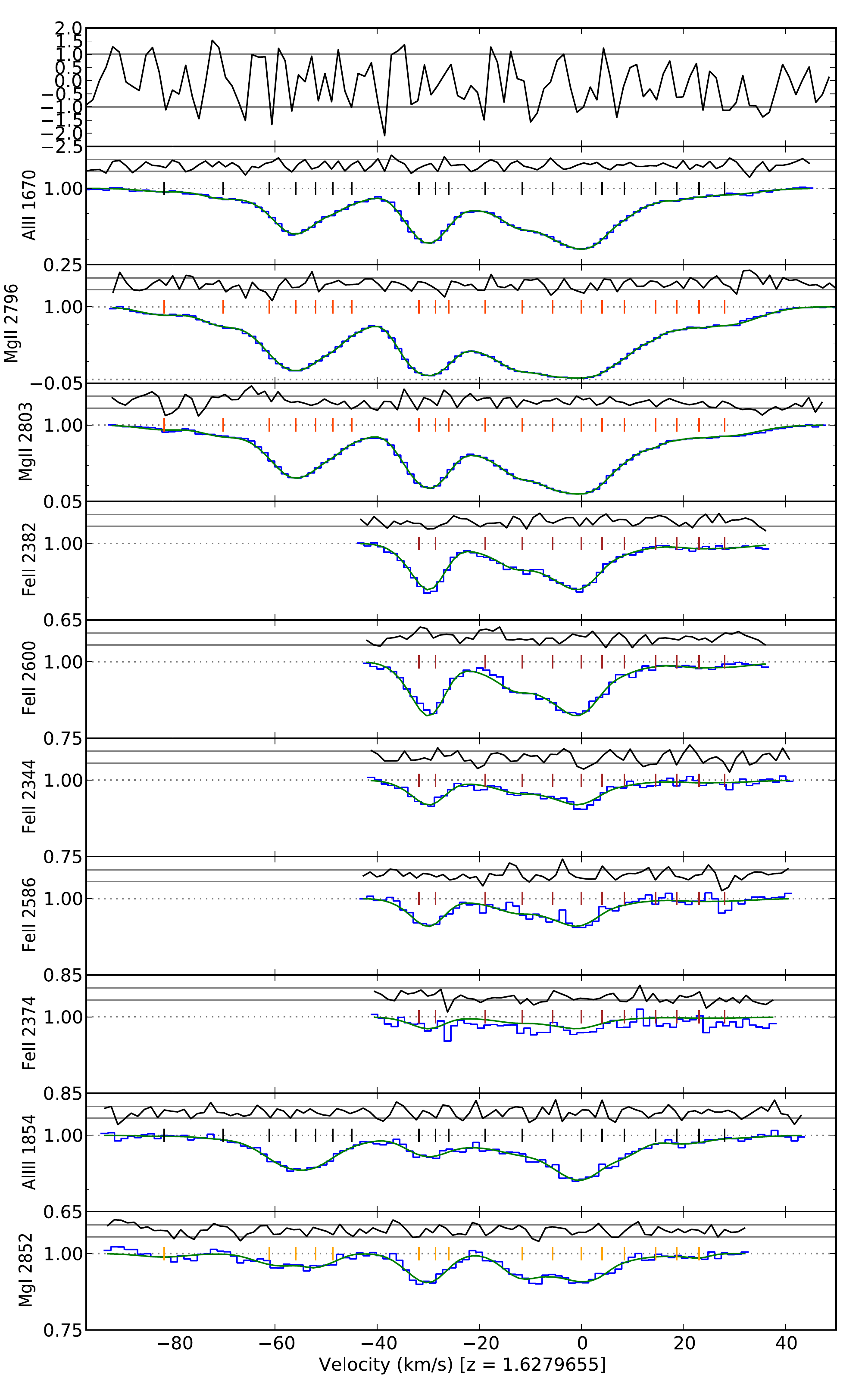}}
  \caption{Absorption profile for system at $\zabs=\systemBRedshift $ with
    transitions used in the second analysis approach for determining
    \daa. The figure is structured as in Fig.~\ref{fig:sys169}.
    Details of the absorption profile fit are given
    in Table~\ref{table:sys162}.}
\label{fig:sys162}
\end{figure}

The transitions used to constrain \daa\ in the second analysis
approach for the absorption system at $\zabs=\systemBRedshift $ are shown in
Fig.~\ref{fig:sys162}. The absorption profile model parameters from
this approach are provided in Table \ref{table:sys162}. The absorption
spans $\sim$100\,\kms\ with three main features which are rather broad
and very asymmetric, revealing a quite complex underlying velocity
structure. Detected transitions not used to constrain \daa\ include
strong \ion{Si}{ii} $\lambda$1526 absorption which falls in the
Ly$\alpha$ forest.

% The absorption is detected in the low-ionization transitions \ion{Al}{ii} $\lambda$1670, the 5 strongest \ion{Fe}{ii} lines between rest-frame wavelengths 2344--2600\,\AA\ \ion{Mg}{ii} $\lambda\lambda$2796, 2803.  A weak neutral magnesium line \ion{Mg}{i} $\lambda$2852 is also present.

% The \ion{Si}{ii} $\lambda\lambda$1260, 1526 are quite strong but since they fall in the Ly$\alpha$ forest are not considered for the \daa analysis.

% Figure sys162

% Table sys162
% \input{newsys162}
\begin{table*}[h!]
\caption{As in Table \ref{table:sys169} but for the absorption system at $\zabs=\systemBRedshift$.}    
\label{table:sys162}      
\centering          
\begin{tabular}{ccccccc cccc ccc}     % 14 columns 
\hline  
  $z$   &  $\sigma_z$ & $b$     & $\sigma_{b}$  & \LogN{Al}{II} & $\sigma_N$  & \LogN{Al}{III}  & $\sigma_N$  & \LogN{Mg}{II} & $\sigma_N$  & \LogN{Mg}{I}  &$\sigma_N$ & \LogN{Fe}{II} & $\sigma_N$  \\
        &             & [\kms]  & [\kms]        &    [\cm]      & [\cm]       & [\cm]           & [\cm]       & [\cm]         & [\cm]       & [\cm]         & [\cm]     & [\cm]         & [\cm] \\          
\hline
1.6272489  &  0.0000071  &  5.7 &  0.9  &  10.69  &  0.10   &  10.30  &  0.47   &  11.48  &  0.07   &  9.96  &  0.20            &         &      \\
1.6273505  &  0.0000094  &  4.9  &  1.3  &  11.13  &  0.14   &  10.79  &  0.26   &  11.81  &  0.15         &         &           &         &      \\
1.6274298  &  0.0000342  &  4.3  &  3.1  &  11.26  &  0.71   &  11.36  &  0.57   &  12.04  &  0.71   &  10.28  &  0.47           &         &      \\
1.6274753  &  0.0000080  &  3.8  &  1.5  &  11.76  &  0.27   &  11.61  &  0.40   &  12.54  &  0.27   &  10.19  &  0.74           &         &      \\
1.6275091  &  0.0000076  &  0.5  &  1.0  &  10.84  &  0.55   &  11.14  &  0.25   &  11.77  &  0.36   &  10.02  &  0.28           &         &      \\
1.6275388  &  0.0000048  &  1.0  &  0.7  &  10.97  &  0.34   &  11.01  &  0.28   &  11.80  &  0.25   &  9.88  &  0.27            &         &      \\
1.6275714  &  0.0000215  &  5.0  &  3.1  &  11.29  &  0.30   &  11.21  &  0.30   &  11.98  &  0.30         &         &           &         &      \\
1.6276866  &  0.0000333  &  4.0  &  1.7  &  11.74  &  1.00   &  11.33  &  1.30   &  12.54  &  0.93   &  10.61  &  0.64     &  11.52  &  0.63      \\
1.6277150  &  0.0000071  &  2.9  &  0.6  &  11.45  &  0.65   &  10.87  &  2.08   &  12.37  &  0.40   &  10.38  &  0.75     &  11.83  &  0.30      \\
1.6277377  &  0.0001969  &  4.9  &  18.5  &  11.34  &  3.06  &  11.15  &  3.07   &  12.05  &  3.07   &  9.57  &  3.04            &         &      \\
1.6278005  &  0.0000141  &  4.3  &  1.8  &  11.29  &  0.26   &  11.15  &  0.26   &  12.15  &  0.21         &         &     &  11.17  &  0.24      \\
1.6278641  &  0.0000061  &  4.3  &  1.0  &  11.74  &  0.12   &  11.45  &  0.14   &  12.61  &  0.13   &  10.74  &  0.09     &  11.72  &  0.12      \\
1.6279165  &  0.0000063  &  2.5  &  1.2  &  11.48  &  0.46   &  11.28  &  0.49   &  12.42  &  0.39   &  10.22  &  0.56     &  11.40  &  0.53      \\
1.6279655  &  0.0000083  &  4.0  &  2.3  &  11.97  &  0.25   &  11.85  &  0.23   &  12.83  &  0.25   &  10.74  &  0.25     &  11.98  &  0.23      \\
1.6280013  &  0.0000079  &  1.2  &  2.0  &  11.11  &  0.74   &  10.54  &  1.92   &  11.91  &  0.85   &  9.96  &  0.69      &  10.86  &  1.35      \\
1.6280395  &  0.0000153  &  3.7  &  2.4  &  11.53  &  0.29   &  11.49  &  0.27   &  12.29  &  0.31   &  10.13  &  0.37     &  11.31  &  0.34      \\
1.6280932  &  0.0000041  &  0.7  &  0.5  &  10.78  &  0.22   &  10.36  &  0.60   &  11.72  &  0.14   &  9.33  &  0.69      &  10.42  &  0.41      \\
1.6281297  &  0.0000036  &  0.8  &  0.6  &  10.82  &  0.06   &  10.80  &  0.16   &  11.51  &  0.07   &  9.42  &  0.55      &  10.27  &  0.36      \\
1.6281677  &  0.0000054  &  0.5  &  1.7  &  10.36  &  0.17   &  10.61  &  0.25   &  11.11  &  0.14   &  9.73  &  0.25      &  10.24  &  0.42      \\
1.6282117  &  0.0000062  &  8.2  &  0.6  &  11.12  &  0.06   &  10.93  &  0.17   &  12.00  &  0.04         &         &     &  11.24  &  0.08      \\

\hline
\daa\ & $\sigma_{\rm stat}$ & $\chi^2_\nu$ &   &&&&   \\
{[ppm]} & {[ppm]}&              &   &&&&  \\
\hline
 \wsystemBdaaTwoPlaces & \wsystemBdaaStatisticalErrorTwoPlaces & \wsystemBdaaChiSquare &&&& \\  % sysb = 1.62
\hline
 \end{tabular}
\end{table*}  

The broad and multi-component nature of the 3 main absorption features
translates to large statistical uncertainties on the redshifts of the
velocity components and, therefore, \daa. The systematic error budget
was determined by conducting the same systematic error tests as
conducted on the $\zabs=\systemARedshift$ absorber in Section
\ref{ssub:systematic_error_tests}. The final result for this
$\zabs=\systemBRedshift $ absorber is then
\begin{equation}\label{eq:res162}
\daa  = \,\ReportStatisticalandSystematicError{\wsystemBdaaOnePlace}
{\wsystemBdaaStatisticalErrorOnePlace}
{\wsystemBdaaSystematicErrorOnePlace}\mbox{\,ppm}\,.
\end{equation}

\subsection{Absorption system at $\zabs=\systemCRedshift$}

\begin{figure}
  \centerline{\includegraphics[width=\columnwidth]{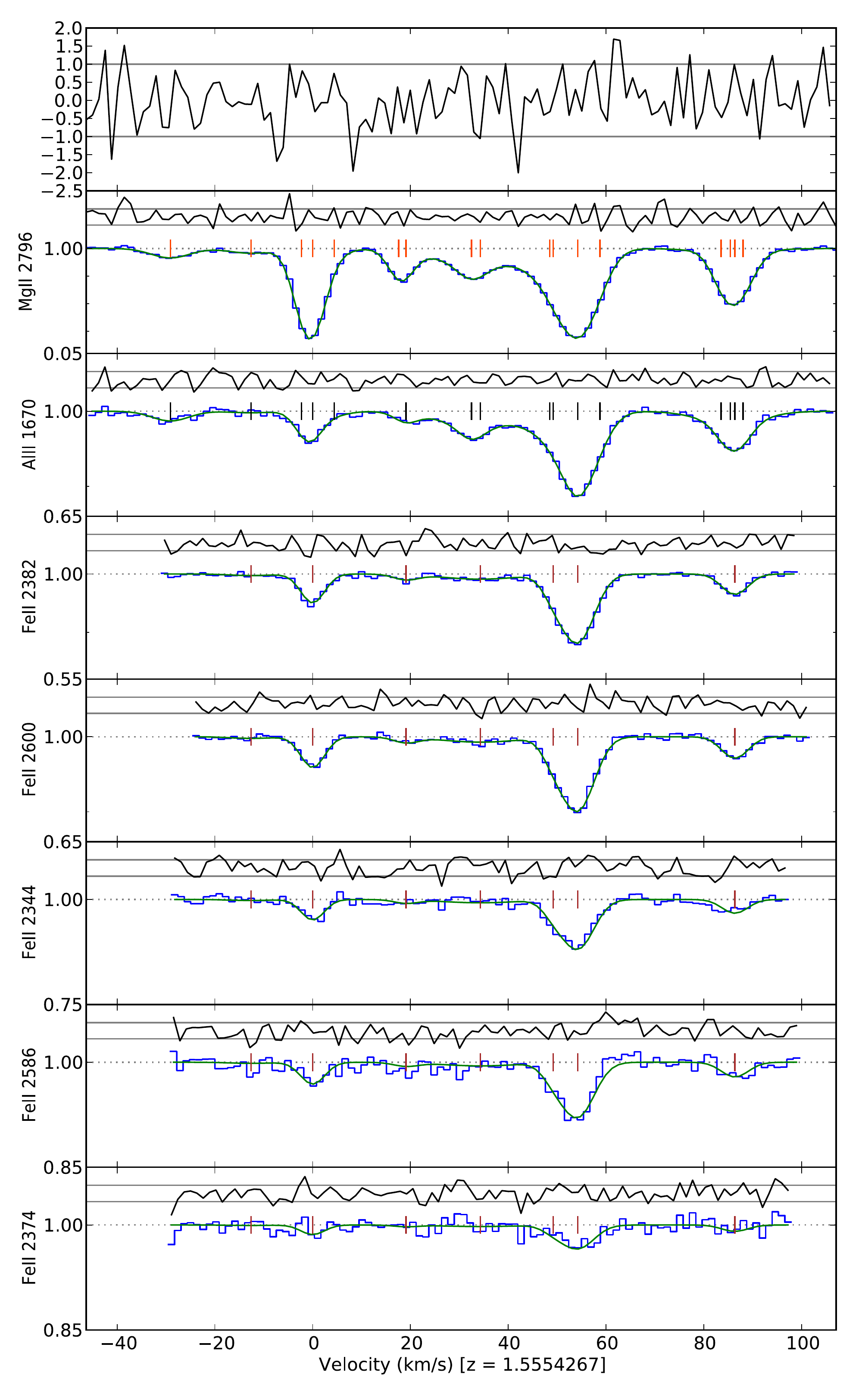}}
  \caption{Absorption profile for system at $\zabs=\systemCRedshift$ with
    transitions used in the second analysis approach for determining
    \daa. The figure is structured as in Fig.~\ref{fig:sys169}.
    Details of the absorption profile fit are given in
    Table~\ref{table:sys155}.}
\label{fig:sys155}
\end{figure}

The transitions used to constrain \daa\ in the second analysis
approach for the absorption system at $\zabs=\systemCRedshift$ are shown in
Fig.~\ref{fig:sys155}. The absorption profile model parameters from
this approach are provided in Table \ref{table:sys155}. The absorption
spans $\sim$100\,\kms\ with three main features which appear at first
to be rather symmetric and narrower than the three features in the
$\zabs=\systemBRedshift$ absorber. Indeed, the underlying velocity structure of
these three features appears to be dominated by a single strong
component in each case. However, our detailed approach to fitting all
the \emph{statistically required} velocity components reveals that
several weaker components are required for an adequate fit. These are
required for the higher optical depth transitions of \ion{Mg}{ii}
$\lambda$2796 and \ion{Al}{ii} $\lambda$1670. There are three telluric
features which contaminate the \ion{Mg}{ii} $\lambda$2803, so this is
excluded from the analysis.

% Figure sys155

% Table sys155
% \input{newsys155}
\begin{table*}[h!]
\caption{As in Table \ref{table:sys169} but for the absorption system at $\zabs=\systemCRedshift$.}    
\label{table:sys155}      
\centering          
\begin{tabular}{cccccccccccc  }     % 7 columns 
%\begin{tabular}{lrrrrrr rrrr}     % 12 columns 
\hline  
  $z$   &  $\sigma_z$ & $b$     & $\sigma_{b}$  & \LogN{Mg}{II} & $\sigma_N$  &\LogN{Al}{II} & $\sigma_N$ & \LogN{Fe}{II} & $\sigma_N$   \\
        &             & [\kms]  & [\kms]        &    [\cm]      & [\cm]       & [\cm]         & [\cm]     & [\cm]         & [\cm]       \\
  \hline
  1.5551786  &  0.0000039  &  4.4  &  0.7  &  11.28  &  0.05          &         &   &  10.61  &  0.18      \\ 
  1.5553193  &  0.0000162  &  5.2  &  2.1 &  11.01  &  0.12    &  10.55  &  0.66   &  9.83  &  1.69      \\ 
  1.5554070  &  0.0000120  &  0.5  &  1.0  &  12.44  &  1.93          &         &   &  10.46  &  0.44      \\ 
  1.5554267  &  0.0000046  &  1.3  &  0.4  &  13.04  &  0.78    &  11.63  &  0.02   &  10.77  &  0.31      \\ 
  1.5554644  &  0.0000496  &  2.8  &  7.8  &  10.86  &  1.14          &         &   &  10.09  &  1.26      \\ 
  1.5555767  &  0.0000074  &  0.5  &  3.1  &  11.63  &  0.78          &         &         &         &      \\ 
  1.5555896  &  0.0000052  &  1.4  &  1.8  &  11.24  &  0.59    &  10.85  &  0.24   &  10.38  &  0.30      \\ 
  1.5557036  &  0.0000026  &  2.1  &  1.1  &  11.37  &  0.16          &         &   &  10.60  &  0.22      \\ 
  1.5557190  &  0.0000094  &  10.7  &  2.5  &  11.89  &  0.08   &  11.34  &  0.08   &  11.12  &  0.09      \\ 
  1.5558400  &  0.0000474  &  4.9  &  3.3 &  11.83  &  0.65          &         &   &  10.99  &  0.65      \\ 
  1.5558459  &  0.0000062  &  0.5  &  1.0  &  11.13  &  0.72    &  11.36  &  0.21   &  9.88  &  2.07      \\ 
  1.5558890  &  0.0000050  &  3.4  &  0.3  &  12.47  &  0.15    &  12.16  &  0.03   &  11.47  &  0.21      \\ 
  1.5559277  &  0.0000161  &  3.0  &  1.4  &  11.45  &  0.38          &         &   &  10.81  &  0.30      \\ 
  1.5561390  &  0.0000372  &  1.1  &  4.6  &  11.55  &  1.35          &         &   &  10.27  &  1.38      \\ 
  1.5561551  &  0.0000132  &  8.3  &  2.6  &  11.15  &  0.49          &         &   &  10.93  &  0.22      \\ 
  1.5561629  &  0.0000047  &  2.1  &  0.4  &  11.74  &  8.58    &  11.49  &  0.03   &  10.75  &  5.19      \\ 
  1.5561771  &  0.0003294  &  2.4  &  16.0  &  11.58  &  11.21        &         &   &  10.39  &  10.88      \\
\hline
\daa\ & $\sigma_{\rm stat}$ & $\chi^2_\nu$ &   &&&&   \\
{[ppm]} & {[ppm]}&              &   &&&&  \\
\hline
  \wsystemCdaaTwoPlaces & \wsystemCdaaStatisticalErrorTwoPlaces & \wsystemCdaaChiSquare &&&& \\  % sysc = 1.55
\hline
 \end{tabular}
\end{table*}  

The $\chi^2$ minimization analysis of the $\zabs=\systemCRedshift$
profiles shown in Fig.~\ref{fig:sys155} and the systematic error tests
conducted in Section \ref{ssub:systematic_error_tests} provide a
final result of
\begin{equation}\label{eq:res155}
\daa  = \,\ReportStatisticalandSystematicError{\wsystemCdaaOnePlace}
{\wsystemCdaaStatisticalErrorOnePlace}
{\wsystemCdaaSystematicErrorOnePlace}\mbox{\,ppm}\,.
\end{equation}
Given the relatively large statistical uncertainty on \daa\ in this
absorber, we neglect here the systematic error budget because it will
be much smaller.

% This system was also studied by \citet{Chand:2004:853} using earlier UVES spectra. They derived constraints on \daa\ only from the stronger spectral feature at velocity $v\approx55$\,\kms in Fig.~\ref{fig:sys155}, reporting $\daa = 2\pm5$\,ppm without including the \ion{Al}{ii} $\lambda$1670 transition. However, using the same spectra and absorption profile model, \citet{Murphy:2007:239001} derived $\daa=1.8\pm6.4$\,ppm with an unacceptably high $\chi^2_\nu$ value of 2.93, ascribing this to data analysis problems in \citet{Chand:2004:853}.

\subsection{Absorption systems at $\zabs=\systemERedshift$ and $\systemDRedshift$}

\begin{figure}
  \centerline{\includegraphics[width=\columnwidth]{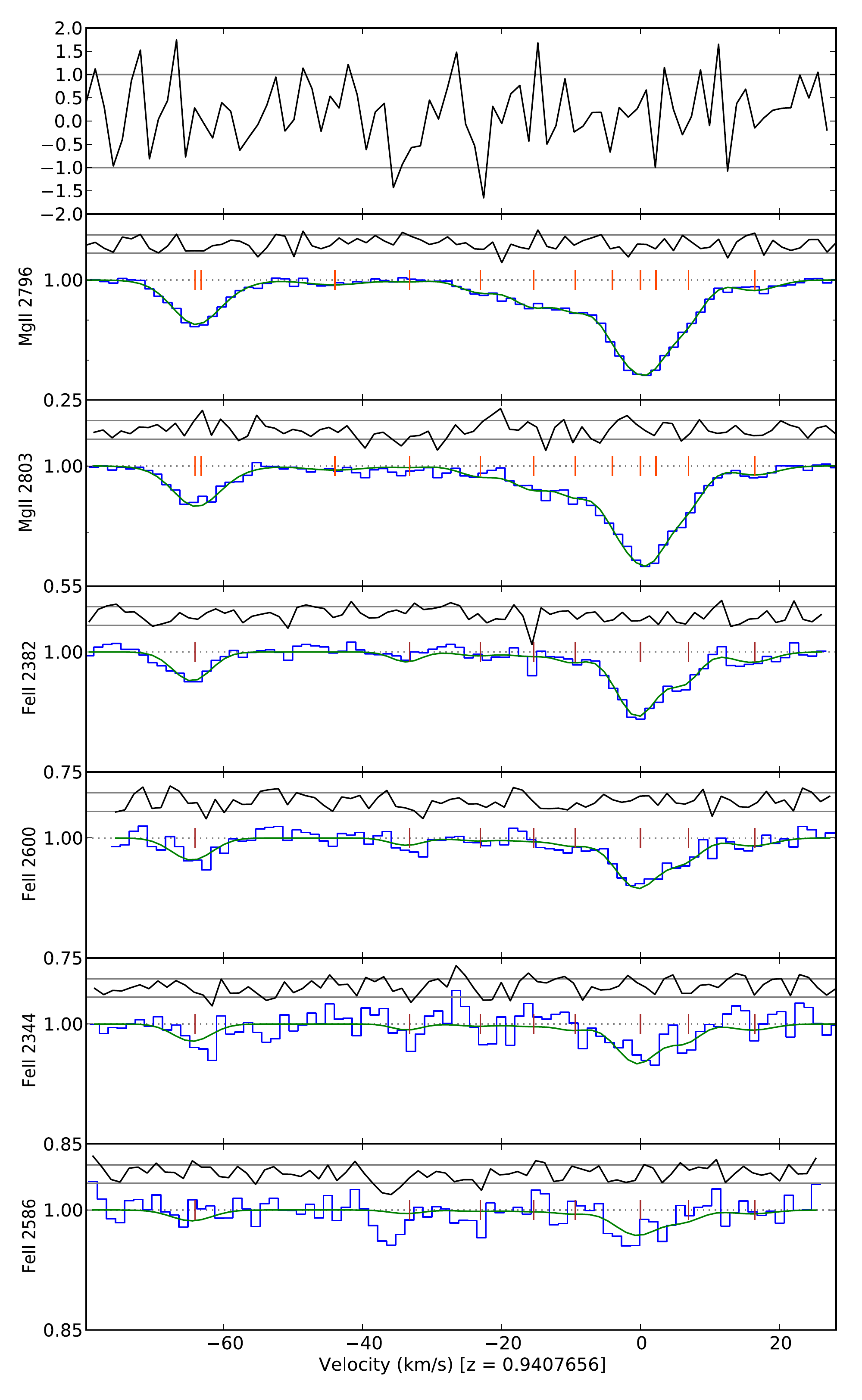}}
  \caption{Absorption profile for system at $\zabs=\systemERedshift$ with
    transitions used in the second analysis approach for determining
    \daa. The figure is structured as in Fig.~\ref{fig:sys169}.
    Details of the absorption profile fit are given
    in Table~\ref{table:sys0940}.}
\label{fig:sys0940}
\end{figure}

\begin{figure}
  \centerline{\includegraphics[width=\columnwidth]{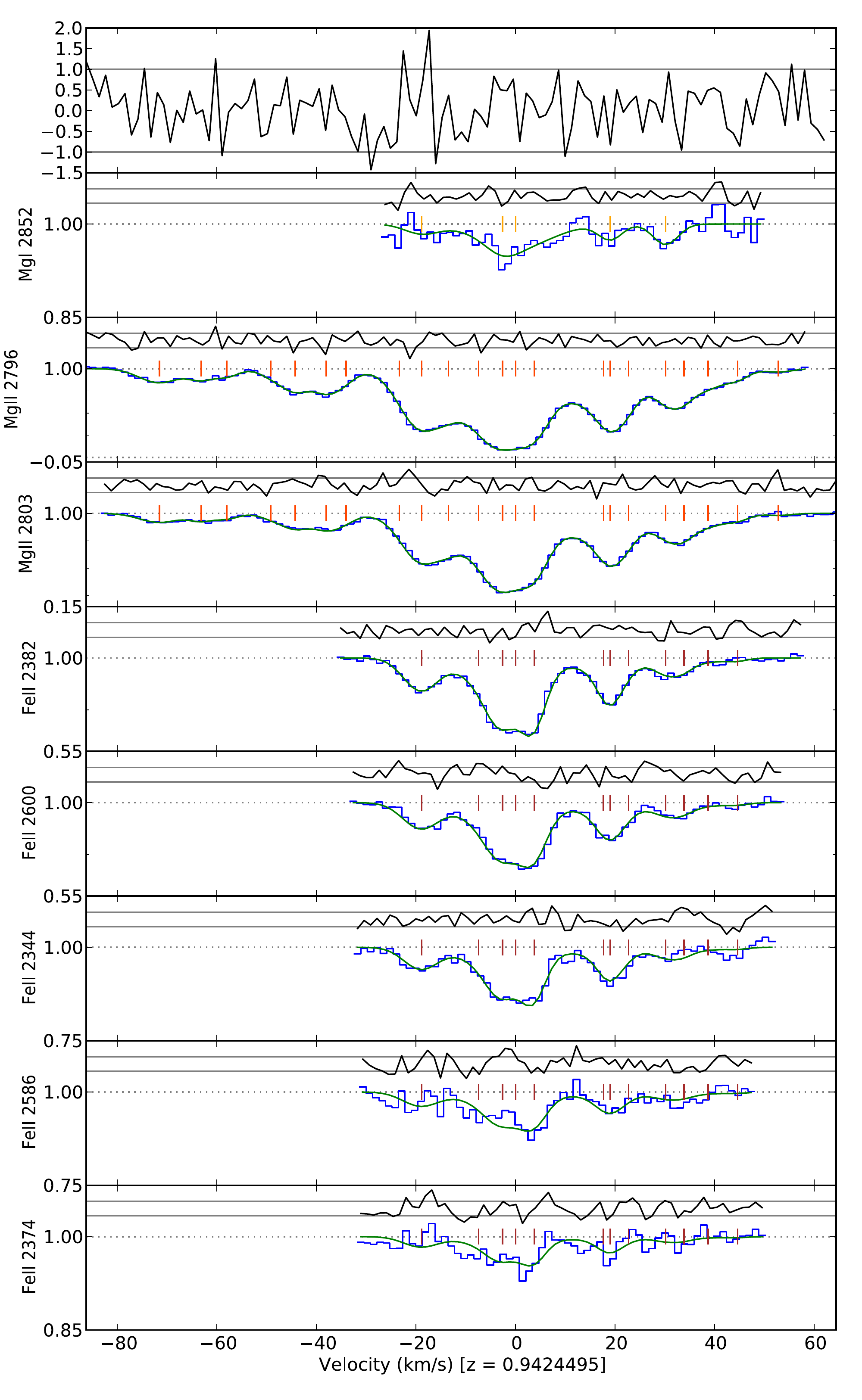}}
  \caption{Absorption profile for system at $\zabs=\systemDRedshift$ with
    transitions used in the second analysis approach for determining
    \daa. The figure is structured as in Fig.~\ref{fig:sys169}.
    Details of the absorption profile fit are given
    in Table~\ref{table:sys0940}.}
\label{fig:sys0942}
\end{figure}

The absorption around $\zabs=0.941$ comprises two absorption
complexes, one with the main absorption at $\zabs=\systemERedshift$ and the
second with the strongest absorption at $\zabs=\systemDRedshift$. They are
separated by $\sim$300\,\kms\ without any hint of absorption in any of
the strongest detected transitions. Hence we fit them separately. The
transitions used to constrain \daa\ in the second analysis approach
for these systems are shown in Figs.~\ref{fig:sys0940} \&
\ref{fig:sys0942}. The absorption profile model parameters from this
approach are provided in Table \ref{table:sys0940}.

% Figure sys0940

% Figure sys0942

% Table sys094
% \input{newsys094}
\begin{table*}[h!]
\caption{As in Table \ref{table:sys169} but for the absorption systems at $\zabs=\systemERedshift$ and $\zabs=\systemDRedshift$.}
% \label{table:sys094}      
\label{table:sys0940}      
\centering          
\begin{tabular}{cccccccccc  }     % 7 columns 

\hline  
  $z$   &  $\sigma_z$ & $b$     & $\sigma_{b}$  & \LogN{Mg}{II}  &  $\sigma_N$     & \LogN{Fe}{II} &$\sigma_N$  & \LogN{Mg}{I} & $\sigma_N$        \\
  &             & [\kms]  & [\kms]        &    [\cm]      & [\cm]       & [\cm]         & [\cm]     & [\cm]         & [\cm]       \\
\hline
0.9419861  &  0.0000024  &  4.5  &  0.7  &  11.58  &  0.03       &         &            &         &      \\
0.9420403  &  0.0000037  &  0.5  &  1.1  &  11.28  &  0.13       &         &            &         &      \\
0.9420738  &  0.0000039  &  0.5  &  1.7  &  11.16  &  0.15       &         &            &         &      \\
0.9421309  &  0.0000122  &  0.5  &  3.2 &  11.07  &  0.44       &         &            &         &      \\
0.9421628  &  0.0000051  &  2.0  &  3.3  &  11.66  &  0.25       &         &            &         &      \\
0.9422031  &  0.0000095  &  1.7  &  3.6  &  11.65  &  0.33       &         &            &         &      \\
0.9422289  &  0.0000151  &  0.5  &  2.2  &  11.25  &  0.58       &         &            &         &      \\
0.9422983  &  0.0000761  &  4.9  &  9.3  &  11.40  &  1.40       &         &            &         &      \\
0.9423275  &  0.0000031  &  3.8  &  0.6  &  12.35  &  0.17 &  11.86  &  0.07      &  9.99  &  0.23      \\ 
0.9423621  &  0.0000040  &  0.5  &  0.5  &  11.73  &  0.15       &         &            &         &      \\
0.9424014  &  0.0000491  &  5.1  &  8.8  &  12.14  &  0.92 &  11.53  &  0.92            &         &      \\
0.9424324  &  0.0000039  &  3.3  &  1.0  &  12.44  &  0.45 &  12.08  &  0.27      &  9.96  &  0.65      \\ 
0.9424495  &  0.0000130  &  9.6  &  2.8  &  12.48  &  0.28 &  11.83  &  0.44      &  10.70  &  0.13      \\
0.9424737  &  0.0000027  &  1.8  &  0.6  &  12.28  &  0.14 &  12.06  &  0.10            &         &      \\
0.9425639  &  0.0000308  &  6.7  &  4.3  &  12.19  &  0.40 &  11.53  &  0.47            &         &      \\
0.9425727  &  0.0000032  &  1.5  &  1.0  &  12.04  &  0.27 &  11.79  &  0.15      &  10.04  &  0.17      \\
0.9425967  &  0.0000107  &  1.6  &  3.6  &  11.62  &  0.80 &  11.01  &  0.80            &         &      \\
0.9426448  &  0.0000048  &  1.9  &  1.7  &  11.81  &  0.26 &  11.36  &  0.21      &  10.19  &  0.13      \\
0.9426687  &  0.0000053  &  0.5  &  0.7  &  11.79  &  0.18 &  11.18  &  0.23            &         &      \\
0.9427000  &  0.0000077  &  2.6  &  3.8  &  11.60  &  0.30 &  10.84  &  0.41            &         &      \\
0.9427385  &  0.0000069  &  0.5  &  1.2  &  11.28  &  0.22 &  10.62  &  0.27            &         &      \\
0.9427913  &  0.0000064  &  2.1  &  2.9  &  10.77  &  0.14       &         &            &         &      \\
\hline % 9
\daa\ & $\sigma_{\rm stat}$ & $\chi^2_\nu$ &   &&&&&   \\
{[ppm]} & {[ppm]}&              &   &&&&&  \\
\hline
\wsystemDdaaTwoPlaces & \wsystemDdaaStatisticalErrorOnePlace & \wsystemDdaaChiSquare &&&&&&&\\
% &&&&&& 16.827 & 12.56 & 1.38\\  
\hline
0.9403508  &  0.0000033  &  2.5  &  0.8  &  11.48  &  0.33 &  11.33  &  0.05  \\ 
0.9403564  &  0.0000052  &  5.0  &  1.6  &  11.55  &  0.27        &          &   \\ 
0.9404810  &  0.0000098  &  4.9  &  3.0  &  10.85  &  0.14        &          &   \\ 
0.9405506  &  0.0000068  &  0.5  &  7.4  &  10.16  &  0.46 &  10.76  &  0.20  \\ 
0.9406165  &  0.0000102  &  2.3  &  2.7  &  11.11  &  0.23 &  10.42  &  0.36  \\ 
0.9406662  &  0.0000043  &  2.4  &  2.7  &  11.47  &  0.17 &  10.49  &  0.39  \\ 
0.9407050  &  0.0000057  &  0.5  &  0.6  &  11.50  &  0.11 &  10.76  &  0.15  \\ 
0.9407394  &  0.0000195  &  1.0  &  3.6  &  11.50  &  0.83        &          &   \\ 
0.9407656  &  0.0000033  &  3.0  &  0.6  &  12.07  &  0.73 &  11.73  &  0.03  \\ 
0.9407801  &  0.0000229  &  1.2  &  5.0  &  11.65  &  1.35        &          &   \\ 
0.9408103  &  0.0000033  &  0.9  &  0.8  &  11.63  &  0.07 &  11.24  &  0.08  \\ 
0.9408722  &  0.0000038  &  3.0  &  1.3  &  11.07  &  0.07 &  10.91  &  0.12  \\
\hline
\daa\ & $\sigma_{\rm stat}$ & $\chi^2_\nu$ &   &&&&&   \\
{[ppm]} & {[ppm]}&              &   &&&&&  \\
\hline
\wsystemEdaaTwoPlaces & \wsystemEdaaStatisticalErrorOnePlace & \wsystemEdaaChiSquare &&&&&&&\\
\hline
 \end{tabular}
\end{table*}

The absorption at $\zabs=\systemERedshift$ spans $\sim$100\,\kms\ and that at
$\zabs=\systemERedshift$ spans $\sim$180\,\kms\ in the strongest transition,
\ion{Mg}{ii} $\lambda$2796. The first system comprises two
well-separated features, while the second comprises four
closely-separated strong features, but in both systems the strong
features are broad and smooth with strong asymmetries. Thus, fairly
complex absorption models were fitted to these systems, with the
strong features comprising several velocity components of similar
strength each.

The $\chi^2$ minimization analysis of the profiles shown in
Figs.~\ref{fig:sys0940} and \ref{fig:sys0942}, and the systematic
error tests in Section \ref{ssub:systematic_error_tests}, provide the
following results:
\begin{equation}\label{eq:res094}
  \daa=\left\{
  \begin{array}{c l}
    \ReportStatisticalandSystematicError{\wsystemEdaaOnePlace}
    {\wsystemEdaaStatisticalErrorOnePlace}{\wsystemEdaaSystematicErrorOnePlace}
    \,{\rm ppm} & \mbox{at } \zabs=\systemERedshift,\\
    \ReportStatisticalandSystematicError{\wsystemDdaaOnePlace}
    {\wsystemDdaaStatisticalErrorOnePlace}{\wsystemDdaaSystematicErrorOnePlace}\,
    {\rm ppm} & \mbox{at } \zabs=\systemDRedshift\,.
  \end{array}
  \right.
\end{equation}
Given the relatively large statistical uncertainties on \daa\ in these
absorbers, we neglect the systematic error budgets because they will
be much smaller.

%This system was studied by Chand \etal (1994) and \citet{Murphy:2007:239001} 
% who found \daa = -12 $\pm$ 7 with $\chi^2_\nu$ =0.9 and 
% \daa = -14.5 $\pm$ 8.5 with $\chi^2_\nu$ =2.43, respectively. 

\subsection{Absorption system at $\zabs=\systemFRedshift$}

\begin{figure}
  \centerline{\includegraphics[width=\columnwidth]{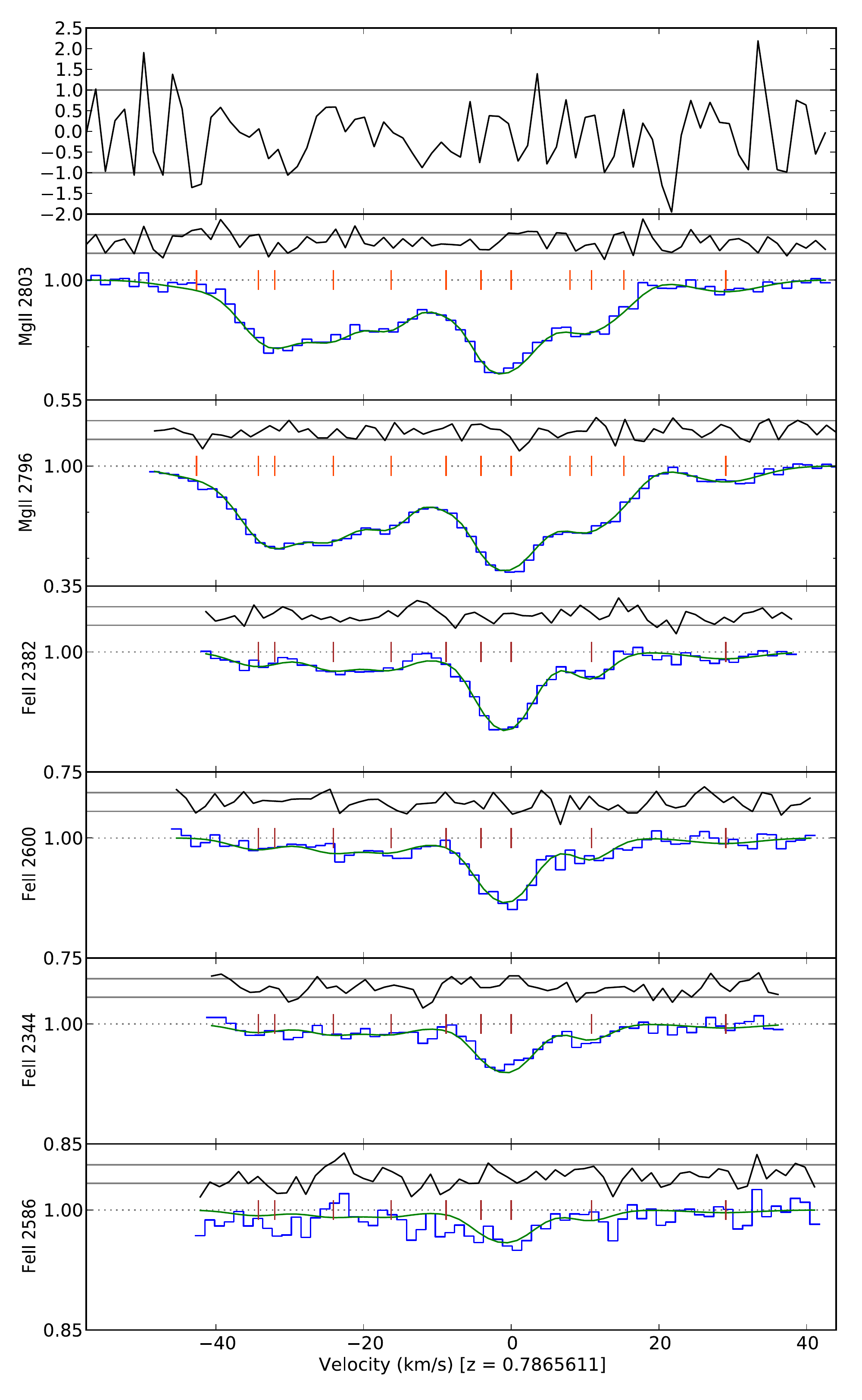}}
  \caption{Absorption profile for system at $\zabs=\systemFRedshift$ with
    transitions used in the second analysis approach for determining
    \daa. The figure is structured as in Fig.~\ref{fig:sys169}.
    Details of the absorption profile fit are given
    in Table~\ref{table:sys078}.}
\label{fig:sys078}
\end{figure}

The transitions used to constrain \daa\ in the second analysis
approach for the absorption system at $\zabs=\systemFRedshift$ are shown in
Fig.~\ref{fig:sys078}. The absorption profile model parameters from
this approach are provided in Table \ref{table:sys078}. The absorption
spans $\sim$90\,\kms\ in one broad complex of several highly blended
features, each of which shows visual evidence of asymmetries and none
of which are particularly narrow. Thus, like all the other absorbers
studied here, the main absorption features require several velocity
components for an adequate fit. Detected transitions not used to
constrain \daa\ include strong \ion{Si}{ii} $\lambda$1526 absorption
which falls in the Ly$\alpha$ forest.

% Figure sys078

% Table sys078
% \input{newsys078}
\begin{table*}[h!]
\caption{As in Table \ref{table:sys169} but for the absorption system at $\zabs=\systemFRedshift$.}
\label{table:sys078}
\centering          
\begin{tabular}{cc ccc ccc }     % 14 columns 
  \hline  
    $z$   &  $\sigma_z$ & $b$     & $\sigma_{b}$  & \LogN{Mg}{II}   &  $\sigma_N$ & \LogN{Fe}{II} &  $\sigma_N$        \\
          &             & [\kms]  & [\kms]        &    [\cm]        & [\cm]       & [\cm]         & [\cm]     \\
\hline
0.7863071  &  0.0000787  &  5.8  &  10.8  &  11.20  &  1.23        &          &   \\ 
0.7863570  &  0.0000483  &  3.7  &  5.5  &  11.21  &  10.38 &  11.11  &  2.40  \\ 
0.7863702  &  0.0000753  &  4.6  &  2.7  &  12.04  &  1.53  &  8.98  &  380.73  \\ 
0.7864176  &  0.0000086  &  3.4  &  4.3 &  11.90  &  0.58  &  11.19  &  0.55  \\ 
0.7864641  &  0.0000091  &  3.5  &  2.9  &  11.87  &  0.30  &  11.18  &  0.30  \\ 
0.7865085  &  0.0000065  &  1.2  &  2.2  &  11.44  &  0.17  &  10.52  &  0.47  \\ 
0.7865367  &  0.0000047  &  0.5  &  0.8  &  11.75  &  0.24  &  11.09  &  0.61  \\ 
0.7865611  &  0.0000067  &  3.7  &  1.2  &  12.14  &  0.18  &  11.85  &  0.12  \\ 
0.7866086  &  0.0001375  &  5.0  &  22.5  &  11.81  &  2.81        &          &   \\ 
0.7866260  &  0.0000052  &  2.3  &  1.0  &  11.46  &  3.37  &  11.29  &  0.06  \\ 
0.7866520  &  0.0000245  &  1.8  &  8.5  &  11.31  &  3.32         &          &   \\ 
0.7867343  &  0.0000040  &  5.6  &  1.1  &  11.36  &  0.05  &  10.90  &  0.11  \\
\hline
\daa\ & $\sigma_{\rm stat}$ & $\chi^2_\nu$ &   &&&&   \\
{[ppm]} & {[ppm]}&              &   &&&&  \\
\hline
   \wsystemFdaaTwoPlaces & \wsystemFdaaStatisticalErrorTwoPlaces & \wsystemFdaaChiSquare &&&& \\  % sysf = 0.78
\hline
 \end{tabular}
\end{table*}  

The $\chi^2$ minimization analysis of the $\zabs=\systemFRedshift$
profiles shown in Fig.~\ref{fig:sys078} and the systematic error tests
in Section \ref{ssub:systematic_error_tests} provide a final result
of
\begin{equation}\label{eq:res078}
\daa =   \ReportStatisticalandSystematicError{\wsystemFdaaOnePlace}
{\wsystemFdaaStatisticalErrorOnePlace}{\wsystemFdaaSystematicErrorOnePlace}
\mbox{\,ppm}\,.
\end{equation}
Given the relatively large statistical uncertainty on \daa\ in this
absorber, we neglect here the systematic error budget because it will
be much smaller.

% Table summary
% \input{newsummary}
\begin{table}
  \caption{Summary of \daa\ results including 1-$\sigma$ statistical uncertainties and estimated systematic errors for the 5 absorption systems towards HE\,2217$-$2818 studied here.
    Our two separate analysis results for the absorber at \zabs=\systemARedshift\ are identified in parentheses.
  }
\label{table:summary}      
\centering
\begin{tabular}{l rcccl}  

\hline
            $z$ & \daa  &   $ \sigma_{\rm stat}$                & $\sigma_{\rm sys}$  & $\chi^2_\nu$ & Ions    \\
            & [ppm]  &   [ppm]                 & [ppm]   &     &      \\
\hline
\systemFRedshift  &  $\wsystemFdaaTwoPlaces$ & \wsystemFdaaStatisticalErrorTwoPlaces & \wsystemFdaaSystematicErrorTwoPlaces & \wsystemFdaaChiSquare & \ion{Fe}{ii},   \ion{Mg}{ii}\\                              % % sysf = 0.78
\systemERedshift  &  $\wsystemEdaaTwoPlaces$ & \wsystemEdaaStatisticalErrorTwoPlaces &  \wsystemEdaaSystematicErrorTwoPlaces & \wsystemEdaaChiSquare & \ion{Fe}{ii},   \ion{Mg}{ii}\\                             % % syse = 0.940
\systemDRedshift  &  $\wsystemDdaaTwoPlaces$ & \wsystemDdaaStatisticalErrorTwoPlaces &  \wsystemDdaaSystematicErrorTwoPlaces & \wsystemDdaaChiSquare & \ion{Fe}{ii},   \ion{Mg}{ii}\\                             % % sysd = 0.942
\systemCRedshift  &  $\wsystemCdaaTwoPlaces$ & \wsystemCdaaStatisticalErrorTwoPlaces & \wsystemCdaaSystematicErrorTwoPlaces & \wsystemCdaaChiSquare & \ion{Fe}{ii},  \ion{Al}{ii}, \ion{Si}{ii}, \ion{Mg}{ii}\\   % % sysc = 1.55
\systemBRedshift  &  $\wsystemBdaaTwoPlaces$ & \wsystemBdaaStatisticalErrorTwoPlaces & \wsystemBdaaSystematicErrorTwoPlaces & \wsystemBdaaChiSquare & \ion{Fe}{ii},  \ion{Al}{ii}, \ion{Si}{ii}, \ion{Mg}{ii, i}\\% % sysb = 1.62
\systemARedshift\ (A)  &  $\psystemAdaaTwoPlaces$ & \psystemAdaaStatisticalErrorTwoPlaces &  & \psystemAdaaChiSquare & \ion{Fe}{ii}, \ion{Al}{ii} \\               % sysa   = 1.69
\systemARedshift\ (A)  & $\psystemAsidamdaaTwoPlaces$ & \psystemAsidamdaaStatisticalErrorTwoPlaces &  & \psystemAsidamdaaChiSquare & \ion{Fe}{ii} \\
\systemARedshift\ (B)  &  $\wsystemAdaaTwoPlaces$ & \wsystemAdaaStatisticalErrorTwoPlaces & \wsystemAdaaSystematicErrorTwoPlaces & \wsystemAdaaChiSquare & \ion{Fe}{ii}, \ion{Al}{ii, iii} \\               % sysa   = 1.69

\hline
 \end{tabular}
\end{table} 

%\begin{figure}
%  \centerline{\includegraphics[width=\columnwidth,angle=-90]{fig_results.pdf}}
%  \caption{Summary of the measurements towards HE\,2217$-$2818}.
%\label{fig:results}
%\end{figure}

\section{Discussion}

Present searches for cosmological variations in the fine-structure
constant are based on the analysis of large samples of absorption
systems observed at different redshifts and directions in the sky.
Thanks to the large size of these samples it is possible to reach an
ensemble sensitivity of a few ppm in \daa. On the other hand, deep
observations of few selected lines of sight are capable of producing
measurements with similar statistical errors. The two approaches have
both intrinsic merits: the former is more effective in detecting
possible changes as a function of redshift or across different regions
of space, and possibly for averaging down systematic effects; the
latter allows more detailed analysis, potentially facilitating the
identification of possible systematic errors which may go undetected
in a large number of systems.

\begin{figure}
  \centerline{\includegraphics[width=\columnwidth]{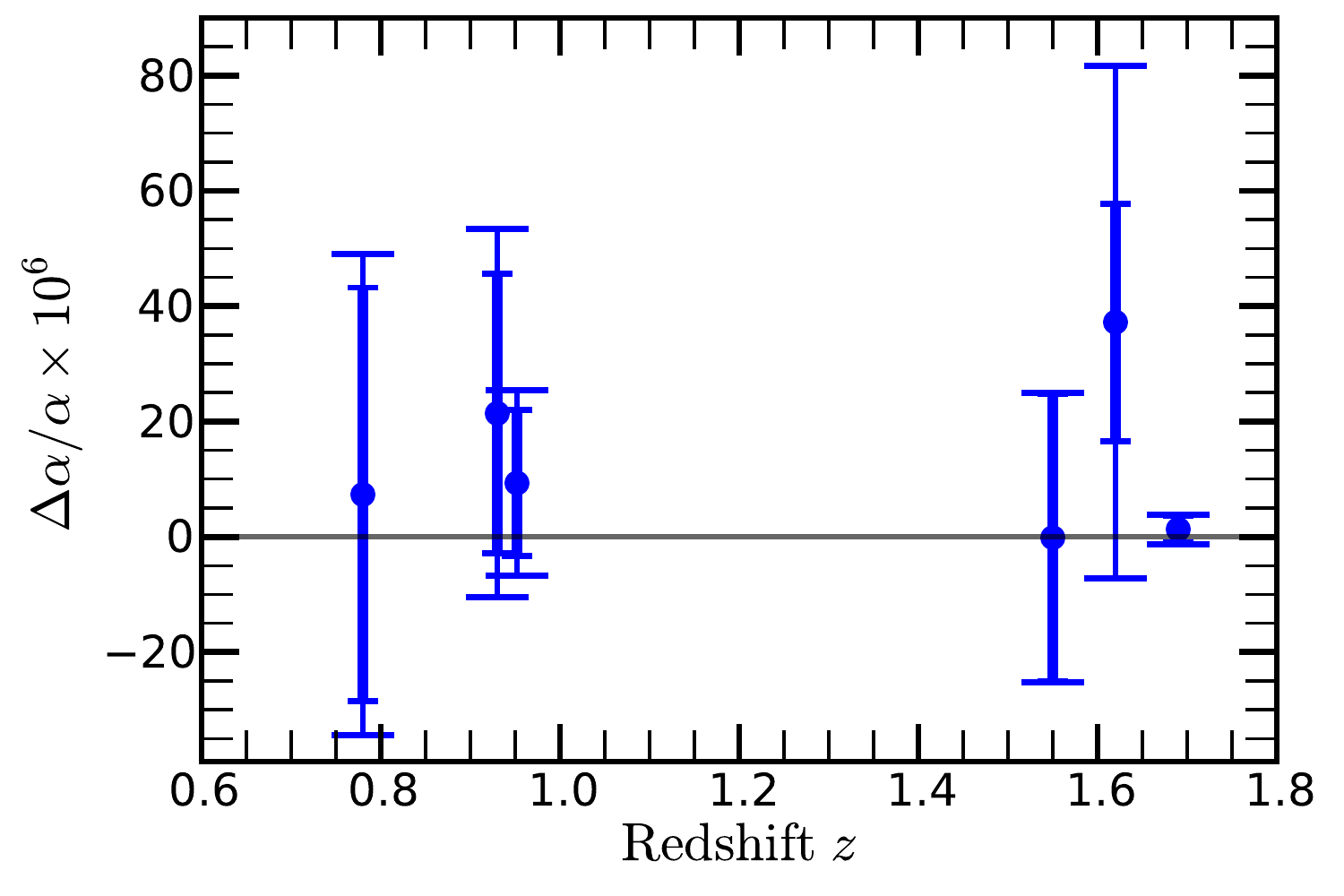}}
  \caption{Summary of \daa\ measurements towards HE\,2217$-$2818.
    Each absorption system's \daa\ value from the second analysis
    approach is plotted with its 1-$\sigma$ statistical uncertainty,
    plus the total uncertainty formed by quadrature addition of the
    statistical and systematic errors in Table \ref{table:summary}.}
\label{fig:summary}
\end{figure}

The line of sight towards the relatively bright quasar HE\,2217$-$2818
is rich with absorption systems. Our Many Multiplet analysis of the 5
systems observed at $\zabs=\systemFRedshift$, \systemERedshift,
\systemDRedshift, \systemCRedshift, \systemBRedshift\ and
\systemARedshift\ provides no evidence of a change of the
fine-structure constant. In particular, the system at
$\zabs=\systemARedshift$ provides a very precise bound on variations
in $\alpha$:
$\daa=\ReportStatisticalandSystematicError{\wsystemAdaaOnePlace}
{\wsystemAdaaStatisticalErrorOnePlace}
{\wsystemAdaaSystematicErrorOnePlace}$\,ppm. This result, stemming
from our second analysis approach in which we conducted several tests
to constrain the systematic error budget, differs somewhat from the
first approach, conducted separately with different details of the
data reduction, calibration and analysis:
$\daa=\ReportStatisticalError{\psystemAdaaOnePlace}
{\psystemAdaaStatisticalErrorOnePlace}$\,ppm. This may indicate that
systematic errors are larger than expected from the range of tests we
conducted. However, it is interesting to note that restricting the
first approach to the 6 \ion{Fe}{ii} transitions, and using a modified
absorption model, yields a similar result to the second approach,
$\daa=\ReportStatisticalError{\psystemAsidamdaaOnePlace}
{\psystemAsidamdaaStatisticalErrorOnePlace}$\,ppm. Clearly, it is
important to continue to refine the analysis techniques to minimize
differences in the results obtained by different approaches. Our
future papers on this and other quasars in our Large Program of new
VLT/UVES observations will aim at that goal.

The main reasons for the small statistical uncertainty in \daa\ from
the $\zabs=\systemARedshift$ absorber are the several narrow absorption features
across its absorption profile, the inclusion of the \ion{Fe}{ii}
$\lambda$1608 transition which, compared to the other \ion{Fe}{ii}
transitions in particular, provide a large range of $q$-coefficients
among the fitted transitions, and the high S/N of the
spectra. Unfortunately, the absorption features characterising the
other 4 absorbers tended to be somewhat broader. And with a smaller
range of $q$ coefficients available from the fitted transitions, the
statistical uncertainties in those absorbers exceeded 10\,ppm.

The bound on \daa\ obtained from the $\zabs=\systemARedshift$ system is one of
the most precise measurements available from an individual absorber.
Three other studies of individual absorbers have claimed measurements
with comparably small uncertainties. These are $\daa=-0.07\pm1.8$\,ppm
obtained from absorption at $\zabs = 1.15$ towards the
extremely bright quasar HE\,0515$-$4414 \citep{Molaro:2008:173},
$\daa=5.4\pm2.5$\,ppm from the system at $\zabs=1.84$ towards quasar
Q\,1101$-$264 \citep{Levshakov:2006:L21}, and $\daa=-1.5\pm2.6$\,ppm from the
system at $\zabs=1.58$ towards HE\,0001$-$2340 \citep{Agafonova:2011:28}. The
former two were both obtained by comparing only \ion{Fe}{ii} lines
with each other, while the latter was derived through a comparison of
\ion{Fe}{ii} and \ion{Si}{ii} transitions. The result at
$\zabs=\systemARedshift$ presented here was obtained by comparing \ion{Al}{ii}
$\lambda$1670 and several \ion{Fe}{ii} transitions, including
\ion{Fe}{ii} $\lambda$1608. Broad agreement is found when using only
the \ion{Fe}{ii} lines, particularly when considering the total
systematic error budget of \wsystemAdaaStatisticalErrorOnePlace\ ppm and 
the strong dependence on
\ion{Fe}{ii} $\lambda$1608 in that case. Therefore, we find no strong
evidence for influence from differential ionization effects which may
be present when comparing different ions.

Recently, \citet{King:2012:3370} analyzed a new sample of 153 measurements
obtained from observations taken with UVES at the ESO Very Large
Telescope (VLT), adding to the previous sample of 143 absorption
systems obtained from the analysis of Keck/HIRES spectra
\citep{Murphy:2003:609}.  The two samples generally probe different directions
in the universe, and rather surprisingly they show different
dependence on redshift; that is, $\alpha$ appears on average to be
smaller in the past in the northern hemisphere but larger in the past
in the southern hemisphere.  This is particularly the case for the
combined dataset at redshift $\ge$1.8. Overall, the variation across
the sky can be simply and consistently represented by a monopole
(i.e.~a constant offset in $\alpha$ from the current laboratory value)
plus a spatial dipole in the direction with right ascension
$17.3\pm1.0$ hours and declination $-61\pm10$ degrees
\citep{King:2012:3370}. The spatial dependence is fitted with
$\daa=m+A\cos\Theta$, where $A$ is the dipole amplitude $A=9.7\pm
2.1$\,ppm, $\Theta$ is the angle on the sky between a quasar
sight-line and the best-fit dipole position, and $m=-1.78\pm0.84$\,ppm
is the monopole term. A simpler model, without the monopole term,
gives $A=10.2\pm2.1$\,ppm with a pole towards right ascension
$17.4\pm0.9$ hours and declination $-58\pm9$ degrees \citep{King:2012:3370}.

The dipole amplitude of $\approx$10\,ppm could in principle be
revealed by precise measurements along individual lines of sight. The
pole and anti-pole directions are observationally difficult regions to
study, particularly the pole which is very close to the Galactic
center and virtually devoid of bright quasars. Towards the directions
for which precise individual absorbers have been studied, the
variation in $\alpha$ expected by the dipole model is small and cannot
be revealed by present accuracies\footnote{The simple dipole model
  used here predicts the following values of \daa\ for the three
  quasars with previous precise measurements: $2.6\pm1.4$\,ppm
  (HE\,0001$-$2340), $2.2\pm1.6$\,ppm (HE\,0515$-$4414) and
  $3.5\pm1.4$ (Q\,1101$-$264).}.  However, the line of sight towards
HE\,2217$-$2818 subtends an angle $\Theta=58.8^\circ$ with respect to
the simple dipole-only model above which is small enough to produce a
signal which may be detectable along this single line of sight: the
predicted signal is $\daa=+5.4\pm1.7$\,ppm. For a simple comparison
with our main result here
($\daa=\ReportStatisticalandSystematicError{\wsystemAdaaOnePlace}
{\wsystemAdaaStatisticalErrorOnePlace}
{\wsystemAdaaSystematicErrorOnePlace}$\,ppm), we combined its
statistical and systematic error terms in quadrature, which reveals
that it differs from this simple dipole prediction (including its
uncertainty) by 1.3\,$\sigma$. The prediction from the monopole+dipole
model at the sky position of HE\,2217$-$2818 is
$\daa=+3.2\pm1.7$\,ppm\footnote{A simple {\sc python} program by
  J.B.W.~for estimating the predicted dipole can be found at
  \url{http://pypi.python.org/pypi/dipole_error/}.}.  Thus, our most
precise result towards HE\,2217$-$2818 does not directly support
evidence for a dipolar variation in $\alpha$ across the sky, though it
does not provide compelling evidence against it either.

\section{Conclusions}

The line of sight towards the relatively bright quasar HE\,2217$-$2818
is rich with absorption systems and is located in the sky at a
favourable position to probe the recently proposed presence of spatial
variation of the fine structure constant, $\alpha$. It is the first
quasar from our ongoing VLT/UVES Large Program to measure \daa\ in
$\sim$25 absorbers towards $\sim$12 bright quasars from which we
expect to make $\sim$20 measurements of \daa\ with precision
$\la$10\,ppm. Approximately 10 of these -- the absorbers which are the
main targets of the Large Program -- will provide very high
statistical precisions of $\sim$2\,ppm each and systematic
uncertainties $\la$2\,ppm. This will yield a final ensemble precision
of better than 1\,ppm with the additional aim of better understanding
possible systematic uncertainties at this accuracy level. Our Many
Multiplet analysis here of the 5 absorption systems observed at
$\zabs=\systemFRedshift$, \systemERedshift, \systemDRedshift, 
\systemCRedshift, \systemBRedshift\  and \systemARedshift\ is
in-keeping with these goals: the highest-redshift absorber provides a
very precise and robust measurement at the $\sim$2\,ppm level while
the 4 others provide weaker constraints $\ga$10\,ppm.

The high S/N, sharp absorption features and diversity of sensitivity
to $\alpha$ in the transitions of the $\zabs=\systemARedshift$
absorber provides the most stringent measurement on \daa\ for this
line of sight. Two separate approaches to the data-reduction,
calibration and absorption line fitting yielded slightly different
results -- $\daa=\ReportStatisticalError {\psystemAdaaOnePlace}
{\psystemAdaaStatisticalErrorOnePlace}$\ ppm and
$\daa=\ReportStatisticalError{\wsystemAdaaOnePlace}
{\wsystemAdaaStatisticalErrorOnePlace}$ -- though both are both
consistent with no variation in $\alpha$. We also investigated a
number of sources of systematic errors in this system and the effect
that they might have on the final measurement. Our final result for
this absorber, including both statistical and systematic error terms,
is $\daa= \ReportStatisticalandSystematicError{\wsystemAdaaOnePlace}
{\wsystemAdaaStatisticalErrorOnePlace}
{\wsystemAdaaSystematicErrorOnePlace}$. The results for the all
absorption systems are summarized in Table~\ref{table:summary}.

If the evidence for a spatial variation of $\alpha$ across the sky
presented by \citet{Webb:2011:191101} and \citet{King:2012:3370} represents reality, then
the simplest model of those results -- a dipolar variation in $\alpha$
-- predicts $\daa=5.4\pm1.7$\,ppm in the direction of the line of
sight of HE\,2217$-$2818. Our constraint on \daa\ from the
$\zabs=\systemARedshift$ absorber does not strongly support or conflict with
that prediction.

\begin{acknowledgements}
 An anonymous referee is acknowledged for several advices.  P.M.~and C.J.M.~acknowledge support from grants PTDC/FIS/111725/2009
  and PTDC/CTE-AST/098604/2008 from FCT, Portugal. The work of
  P.M.~was partially funded by grant CAUP-06/2010-BCC. The work of
  C.J.M.~is funded by a Ci\^encia 2007 Research Contract, funded by
  FCT/MCTES (Portugal) and POPH/FSE (EC). M.T.M.~thanks the Australian
  Research Council for funding under the {\it Discovery Projects}
  scheme (DP110100866). PPJ is supported in part by Agence Nationale pour
  la Recherche under contract ANR-10-Blan-0510-01.
  RS and PPJ gratefully acknowledge support
from the Indo-French Centre for the Promotion of Advanced Research
(Centre Franco-Indien pour la promotion de la recherche
avance) under Project N.4304-2. SL has been supported by FONDECYT grant number 1100214.
The work of I.I.A. and S.A.L. is supported by DFG Sonderforschungsbereich
SFB 676 Teilprojekt C4 and, in part, by the State Program ÔLeading
Scientific Schools of Russian FederationÕ (grant NSh 4035.2012.2)
\end{acknowledgements}

\end{document}